\documentclass[journal]{IEEEtran}

\usepackage{graphicx} 
\DeclareGraphicsExtensions{.eps}
  
\usepackage{subcaption} 
  
\usepackage{tikz}
\usepackage{pgfplots,pgfplotstable}
\usetikzlibrary{pgfplots.groupplots}
\pgfplotsset{compat=1.5}

\usepackage{multirow}

\usepackage{amsmath}

\usepackage{textcomp}
\usetikzlibrary{shapes,arrows}

\usepackage{empheq} 

\usepackage{stfloats} 

\usepackage[hyphens]{url}
\usepackage[hidelinks]{hyperref}
\hypersetup{colorlinks=true,linkcolor=black,citecolor=black,filecolor=black,urlcolor=black,breaklinks=true}
\usepackage{cite} 

\usepackage{color}

\newcommand{\norm}[1]{\left\lVert#1\right\rVert} 
\usepackage{cancel} 

\usepackage{array} 

\usepackage[export]{adjustbox} 

\usepackage[normalem]{ulem} 

\usepackage{amsfonts}

\makeatletter
\def\bstctlcite{\@ifnextchar[{\@bstctlcite}{\@bstctlcite[@auxout]}}
\def\@bstctlcite[#1]#2{\@bsphack
 \@for\@citeb:=#2\do{%
   \edef\@citeb{\expandafter\@firstofone\@citeb}%
   \if@filesw\immediate\write\csname #1\endcsname{\string\citation{\@citeb}}\fi}%
 \@esphack}
\makeatother

\begin{document}

\bstctlcite{IEEEexample:BSTcontrol}

\title{Assigning Shadow Prices to \\ Synthetic Inertia and Frequency Response Reserves from Renewable Energy Sources}

\author{Luis~Badesa,~\IEEEmembership{Member,~IEEE},
        Carlos~Matamala,~\IEEEmembership{Student Member,~IEEE},
        Yujing~Zhou,
        and~Goran~Strbac,~\IEEEmembership{Member,~IEEE}
\thanks{
L. Badesa, C. Matamala and G. Strbac are with the Department of Electrical and Electronic Engineering, Imperial College London, SW7 2AZ London, U.K. (emails: \{luis.badesa, c.matamala-vergara, g.strbac\}@imperial.ac.uk). Y. Zhou was with the Department of Electrical and Electronic Engineering, Imperial College London.}
\thanks{
L. Badesa and G. Strbac were supported by UKERC project `Market and Policy Design for Ambitious Wind Generation', and EU H2020 project TradeRES (grant 864276). C. Matamala was supported by the National Agency for Research and Development (ANID) through scholarship ANID/PROGRAMA BECAS CHILE DOCTORADO 2020-72210414.}
\vspace{-1em}
}

\markboth{IEEE Transactions on Sustainable Energy, August~2022}
{Shell \MakeLowercase{\textit{et al.}}: Bare Demo of IEEEtran.cls for IEEE Journals}

\maketitle

\begin{abstract}
Modern electricity grids throughout the world, particularly in islands such as Great Britain, face a major problem on the road to decarbonisation: the significantly reduced level of system inertia due to integration of Renewable Energy Sources (RES). Given that most RES such as wind and solar are decoupled from the grid through power electronics converters, they do not naturally contribute to system inertia. However, RES could support grid stability through appropriately controlling the converters, but currently no market incentives exist for RES to provide this support. In this paper we develop a methodology to optimally clear a market of ancillary services for frequency control, while explicitly considering the participation of grid-forming and grid-following inverter-based technologies. We propose a mathematical framework that allows to compute shadow prices for ancillary services offered by a pool of diverse providers: synchronous and synthetic inertia, enhanced frequency response (e.g.~from curtailed RES) and traditional primary frequency response (e.g.~by thermal generators). Several case studies are run on a simplified Great Britain system, to illustrate the applicability and benefits of this pricing scheme.
\end{abstract}

\begin{IEEEkeywords}
Shadow pricing, frequency stability, renewable energy, synthetic inertia.
\end{IEEEkeywords}

\IEEEpeerreviewmaketitle

\section*{Nomenclature}
\addcontentsline{toc}{section}{Nomenclature}

\vspace*{-2pt}
\subsection*{Indices and Sets}
\begin{IEEEdescription}[\IEEEusemathlabelsep\IEEEsetlabelwidth{$\textrm{RoCo}$}]
    \item[$i$] Index of inverter-based resources.
    \item[$\mathcal{I}$] Set of inverter-based resources providing frequency support.
    \item[$g,\,\, \mathcal{G}$] Index, Set of generators.
\end{IEEEdescription}

\vspace*{-6pt}
\subsection*{Constants and Parameters}
\begin{IEEEdescription}[\IEEEusemathlabelsep\IEEEsetlabelwidth{$\Delta f_{\textrm{ma}}$}]
    \item[$\alpha$] Wind forecast error in the day-ahead, as percentage of installed capacity.
    \item[$\Delta f_{\textrm{max}}$] Maximum admissible frequency deviation at the nadir (Hz).
    \item[$\mathrm{c}_{g}^{\mathrm{st}}$] Start-up cost of generator $g$ (\pounds).
    \item[$\textrm{c}^{\textrm{m}}_g$] Marginal cost of generator $g$ (\pounds/MWh).
    \item[$\textrm{c}^{\textrm{nl}}_g$] No-load cost of generator $g$ (\pounds/h).
    \item[$f_0$] Nominal frequency of the power grid (Hz).
    \item[$\textrm{H}_g$] Inertia constant of generator $g$ (s).
    \item[$\textrm{H}_i$] Synthetic inertia constant of inverter-based resource $i$ (s).
    \item[$\textrm{k}_\textrm{rec}$] Recovery factor for synthetic inertia ($\textrm{s}^{-1}$).
    \vspace{0.5mm}
    \item[$\textrm{P}_\textrm{D}$] Total system demand (MW).
    \vspace{0.5mm}
    \item[$\textrm{P}^\textrm{max}_g$] Rated power of generator $g$ (MW).
    \vspace{0.5mm}
    \item[$\textrm{P}^\textrm{msg}_g$] Minimum stable generation of generator $g$ (MW).
    \vspace{0.5mm}
    \item[$\textrm{P}_i$] Power available from inverter-based resource $i$ (MW).
    \item[$\textrm{P}^\textrm{max}_i$] Rated power of inverter-based resource $i$ (MW).
    \vspace{0.5mm}
    \item[$\textrm{R}^\textrm{max}_g$] Frequency response capacity of generator $g$ (MW).
    \item[$\textrm{R}^\textrm{max}_i$] Frequency response capacity of inverter-based resource $i$ (MW).
    \vspace{0.5mm}
    \item[$\textrm{RoCoF}_\textrm{max}$] \qquad Maximum admissible RoCoF (Hz/s).
    \item[$\textrm{T}_\textrm{EFR}$] Delivery time of EFR service (s).
    \item[$\textrm{T}_\textrm{PFR}$] Delivery time of PFR service (s).
    \vspace{0.5mm}
    \item[$\textrm{T}_g^\textrm{mdt}$] Minimum down time for generator $g$ (h).
    \vspace{0.5mm}
    \item[$\textrm{T}_g^\textrm{mut}$] Minimum up time for generator $g$ (h).
    \vspace{0.5mm}
    \item[$\textrm{T}_g^\textrm{st}$] Start-up time for generator $g$ (h).
\end{IEEEdescription}

\vspace*{-6pt}
\subsection*{Decision Variables}
\begin{IEEEdescription}[\IEEEusemathlabelsep\IEEEsetlabelwidth{$\lambda_1, \lambda_2$}]
    \item[$\lambda_1, \lambda_2, \mu$] \quad Dual variables for the SOC nadir constraint.
    \item[$\lambda_\textrm{q-s-s}$] Dual variable for the q-s-s constraint.
    \item[$\lambda_\textrm{RoCoF}$] Dual variable for the RoCoF constraint.
    \item[$P_g$] Power produced by generator $g$ (MW).
    \item[$P_\textrm{L}$] Largest power infeed (MW).
    \item[$P^\textrm{curt}_i$] Power curtailed from inverter-based resource $i$ (MW).
    \item[$R_i$] Headroom in inverter-based resource $i$ for providing EFR (MW).
    \item[$R_g$] Headroom in generator $g$ for providing PFR (MW).
    \item[$y_g$] Binary variable, commitment state of generator $g$.
    \item[$y_g^\textrm{sd}$] Shut-down variable for generator $g$.
    \vspace{0.5mm}
    \item[$y_g^\textrm{sg}$] `Start generating' state of generator $g$.
    \vspace{0.5mm}
    \item[$y_g^\textrm{st}$] Start-up variable for generator $g$.
\end{IEEEdescription}

\vspace*{-6pt}
\subsection*{Linear Expressions \normalfont{(linear combinations of decision variables)}}
\begin{IEEEdescription}[\IEEEusemathlabelsep\IEEEsetlabelwidth{$\lambda_1, \lambda_2$}]
    \item[$H_\textrm{sync}$] Aggregate synchronous inertia from thermal generators (MW$\cdot \textrm{s}$).
    \item[$H_\textrm{synt}$] Aggregate synthetic inertia from GFM inverters (MW$\cdot \textrm{s}$).
    \item[$R_\mathcal{I}$] Aggregate headroom in inverter-based resources for providing EFR (MW).
    \item[$R_\mathcal{G}$] Aggregate headroom in generators for providing PFR (MW).
\end{IEEEdescription}

\section{Introduction}

\IEEEPARstart{O}{ne} of the main challenges in operating power grids with high penetration of Renewable Energy Sources (RES) is maintaining frequency stability following a sudden imbalance between generation and demand. This is due to the non-synchronous nature of the dominant RES in most countries, i.e.~wind turbines and solar photovoltaic panels, which do not contribute to system inertia therefore leading to faster frequency variations. A recent event highlighting the challenge in operating low-inertia systems was the national lockdown in Great Britain driven by the COVID-19 pandemic in 2020, when the cost of ancillary services for frequency control increased three times compared to the previous year, due to the high percentage of demand covered by zero-carbon sources during this period \cite{LuisCovid}.

In the near future, RES are expected to largely replace synchronous generators in most systems to meet decarbonisation targets. In fact, inverter-based resources have the capability to contain frequency variations if appropriately controlled. Two main categories of grid-connected inverters have been defined, depending on their controls: grid-following (GFL) and grid-forming (GFM) technologies. GFM inverters are a widely discussed option as they would allow a grid to operate without any synchronous machine \cite{NREL_GFM}. However, virtually all of the currently installed capacity of RES in the world is based on GFL control, which could also provide some level of frequency support when operating in combination with either synchronous machines or GFM devices \cite{Matevosyan_PES1}.

Extensive research has been conducted to investigate the capability of wind turbines to provide frequency support \cite{Wu2018State-of-the-artSystems}. 
GFL inverters with a droop control to regulate power output as a function of frequency deviation can mimic the governor control of synchronous generators \cite{Ela_NREL}. The frequency response from inverters can in fact be significantly faster than that of synchronous machines, since the latter are subject to the slow dynamics of their electromechanical elements coupled to the grid  \cite{Matevosyan_PES2}. However, a key feature of GFL control is the Phase-Locked Loop (PLL) used for tracking the voltage phase angle of the grid and then synchronising to it, which is also used to estimate frequency. The PLL has an inherent measurement delay, therefore GFL cannot provide frequency support at the very instant when the frequency decline starts, which prevents them for providing an inertia-like service equivalent to synchronous inertia \cite{Paolone_PSCC}.

GFM inverters do not require a PLL, therefore they can provide frequency support at the very instant of a contingency, in a similar manner to a synchronous generator \cite{Paolone_PSCC}. These inverters behave as voltage sources that impose frequency to the grid, instead of injecting current with a frequency equal to the one measured from the grid through a PLL, as GFL inverters do. Therefore, GFM inverters do not have any delay in responding to grid contingencies. Although there is no clear definition for the service of `synthetic inertia', here we use this term for an injection of power proportional to the derivative of frequency and initiated exactly at the start of a generation/demand outage, that is, an equivalent service to synchronous inertia. In the case of synthetic inertia provided by wind turbines, the kinetic energy stored in their rotating masses can be used for this service, which has the effect of slowing down the turbine therefore deviating it from the maximum power point \cite{GE_AustraliaReport}. The wind turbine can deliver extra power for a few seconds by extracting this kinetic energy, while it must later go back to optimal rotating speed by delivering less power to the grid than it was delivering pre-fault. This `recovery effect' is essentially a smaller power imbalance that occurs some time after the initial contingency.

Although inverter-based RES already have the technical capability to largely replace synchronous generators in performing frequency control, currently no market incentives exists for RES to provide this type of ancillary services. Technology manufacturers therefore have no clear incentive to deploy GFM technology, as there is no clarity on how the services provided by GFM inverters would be rewarded. There is a gap in the literature in designing appropriate economic incentives for these technologies, assigning prices that correspond to the value they bring to system operation. Previously proposed pricing schemes for frequency-control ancillary services do not explicitly distinguish between GFL and GFM capabilities. References \cite{ElaI,ERCOT_EFR} propose linear approximations of the frequency-security conditions to obtain shadow prices for ancillary services. The work in \cite{PaturetPricing} focuses on pricing inertia from its impact on the Rate-of-Change-of-Frequency (RoCoF) constraint, while the frequency nadir has been show to be the binding requirement in highly renewable systems.
Similarly, reference \cite{DvorkinPricing} has introduced a stochastic formulation for pricing of inertia, while only focusing on the RoCoF requirement. 

The most relevant previous work is \cite{ManuelGarciaHICSS}, where the authors recently proposed a method to price fast frequency response reserve and synthetic inertia from inverter-based resources. Following the service definitions by the Texas system operator, ERCOT, fast response is modelled as a step function, and a requirement to sustain the power injection during 15min is enforced. No distinction between synthetic inertia from GFL and GFM sources is made, however an implicit assumption on all inverters acting as GFL is used, since synthetic inertia is assumed to have a small delay. In addition, synchronous inertia cannot be considered as a decision variable, and the overall formulation is non-convex due to the nadir constraint, therefore requiring an iterative algorithm to converge to a Karush-Kuhn-Tucker point and then compute shadow prices for ancillary services. Furthermore, some important characteristics of RES are not accounted for in any of the above works, such as the recovery effect of wind turbines, which is critically important as it increases the need for frequency response that can be sustained in time \cite{FeiAssessment}. 

In the present article, a methodology for pricing ancillary services from RES is proposed, based on the frequency-secured scheduling of a power system. The formulation introduced in \cite{LuisPricing} is enhanced to explicitly consider the frequency-support capabilities of inverter-based resources with non-dispatchable energy sources, e.g.~wind turbines and solar PV. A market design based on this methodology would create a revenue stream for RES owners, with the corresponding prices leading to appropriate investment signals. The contributions of this paper are as follows:
\begin{enumerate}
    \item To propose a novel mathematical formulation to assign shadow prices for ancillary services from RES, while considering the capabilities of grid-following and grid-forming inverters.
    \item To explicitly distinguish, for the first time, the specific characteristics of RES providing frequency support within the pricing scheme, such as the recovery effect associated with synthetic inertia. A method for optimising the synthetic inertia constant for GFM units is introduced.
    \item To provide new insight on the incentives for RES to provide ancillary services under the proposed pricing framework, and the implications of different methodologies for dealing with non-convexities such as dispatchable, restricted, and convex hull pricing.
\end{enumerate}

The remainder of this paper is structured in the following manner: Section~\ref{sec:Methodology} describes the mathematical methodology used to compute shadow prices for ancillary services in any given system. Section~\ref{sec:CaseStudies} illustrates the applicability of said methodology through relevant case studies representative of low-carbon power systems, including a discussion on its compatibility with different pricing approaches for dealing with non-convexities. Finally, Section~\ref{sec:Conclusion} gives the conclusion.

\section{Methodology for shadow pricing \\of ancillary services} \label{sec:Methodology}

This sections discusses the proposed framework for a frequency-secured scheduling of a power system, from which shadow prices for ancillary services are derived. The deductions included here make no assumption on the particular generation mix of a system, and therefore are applicable to any power grid under consideration.

\begin{figure}[t!]
    \centering
    \includegraphics[width=0.9\linewidth]{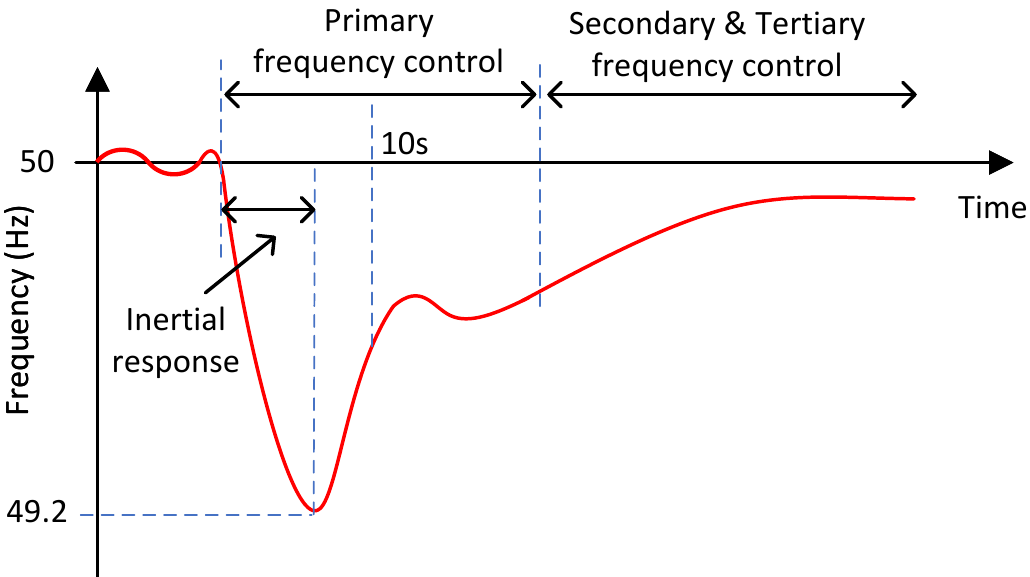}
    \caption{Schematic of system frequency following an incident, along with the different frequency control services. The value of `10s' for primary frequency control corresponds to the requirement in Great Britain, and is used simply for illustrative purposes. Similarly, the value of 49.2Hz is the lowest admissible frequency in GB.}
    \label{fig:freq_drop}
\end{figure}

Ancillary services for frequency control are scheduled to contain a frequency drop ensuing the loss of a generation unit, depicted in Fig.~\ref{fig:freq_drop}. After a sudden and unexpected power loss, the kinetic energy stored in the rotating masses of synchronous generators is spontaneously released to compensate the power imbalance, with the effect of slowing down the rotating speed of the generators and therefore decreasing system frequency. The drop in frequency activates the governor controls of generators operating with some headroom, which then increase their fuel injection to increase power output and help compensate the power imbalance \cite{KundurBook}. This service is referred to in this paper as Primary Frequency Response (PFR), delivered by electromechanical devices such as synchronous generators. A similar but faster service is the Enhanced Frequency Response (EFR), delivered by power-electronics based devices such as wind turbines operating with some curtailment, that can provide a faster power injection in the order of one second after the power imbalance starts \cite{LuisCovid}.

The frequency nadir (lowest value of system frequency during a transient) occurs in a timescale of several seconds. A second nadir may occur as a consequence of the `recovery effect' of wind turbines providing synthetic inertia, while this effect is compensated by the frequency response services as explained in Section~\ref{sec:SwingModel}. In a longer timescale, the secondary and tertiary frequency controls eventually take the system frequency back to its nominal value. As secondary and tertiary controls are not affected by the level of inertia in the system (which only affects the frequency performance in the first few seconds after an outage), previous works have already analysed their role in generation scheduling schemes \cite{GoranBook}.

In summary, four ancillary services for frequency control are considered in this paper: synchronous inertia, synthetic inertia, EFR (power injection fully delivered by 1s after a contingency, i.e. all headroom must be exhausted no later than 1s following the contingency) and PFR (power injection fully delivered by 10s after a contingency). Note that these values of `$\textrm{T}_\textrm{EFR}=1\textrm{s}$' and `$\textrm{T}_\textrm{PFR}=10\textrm{s}$' correspond to the service definitions in Great Britain \cite{NationalGridRequirements}, however the optimisation framework proposed in this paper can accommodate a different definition of speed for these services, by simply changing the values of these two parameters. 

Without loss of generality, we consider wind as the only RES. While wind has kinetic energy stored in the turbine that can be used for provision of synthetic inertia, solar PV could provide synthetic inertia if equipped with a GFM converter and some additional energy storage system, such as an ultracapacitor bank, as demonstrated in the EASY-RES project \cite{EasyRES}. Battery storage systems with GFM inverters are another example of a converter-interfaced asset that could provide synthetic inertia with no recovery.

\subsection{Swing-equation model for system frequency} \label{sec:SwingModel}

The electric frequency of a power system must always be kept within narrow limits around the nominal operating point (e.g.~50Hz in European grids) to avoid any disruption to energy customers. System frequency can be mathematically described through the swing equation \cite{KundurBook}:
\begin{equation} \label{eq:SwingEq}
    \frac{2(H_\textrm{sync} + H_\textrm{synt})}{f_0}\cdot\frac{\textrm{d} \Delta f(t)}{\textrm{d} t} 
    = 
    \textrm{FR}(t) - P_\textrm{L} - \textrm{P}_\textrm{rec}(t)
\end{equation}
Where a positive value for $P_\textrm{L}$ implies a loss of generation. Term $\textrm{FR}(t)$ is the aggregate frequency response reserve in the system, composed by the contributions from power-electronics devices (with fast dynamics, able to provide the EFR service) and synchronous generators (slower due to their mechanical elements, therefore providing PFR):
\begin{equation} 
\textrm{FR}(t) = \textrm{EFR}(t) + \textrm{PFR}(t)
\end{equation}
\begin{equation} \label{eq:EFRdefinition}
    \textrm{EFR}(t)=\left\{ 
    \begin{array}{ll}
         \sum_{ i \in \mathcal{I} } R_i \cdot \frac{ t } { \textrm{T}_\textrm{EFR} } \quad & \mbox{if $t \leq \textrm{T}_\textrm{EFR}$} \\
        \sum_{ i \in \mathcal{I} } R_i \quad & \mbox{if $ t > \textrm{T}_\textrm{EFR} $}
    \end{array}
    \right.
\end{equation}
\begin{equation} \label{eq:PFRdefinition}
    \textrm{PFR}(t)=\left\{ 
    \begin{array}{ll}
         \sum_{ g \in \mathcal{G} } R_g \cdot \frac{ t } { \textrm{T}_\textrm{PFR} } \quad & \mbox{if $t \leq \textrm{T}_\textrm{PFR}$} \\
        \sum_{ g \in \mathcal{G} } R_g \quad & \mbox{if $ t > \textrm{T}_\textrm{PFR} $}
    \end{array}
    \right.
\end{equation}
The droop characteristics of frequency control are approximated in~(\ref{eq:EFRdefinition}) and~(\ref{eq:PFRdefinition}) by a linear function, a procedure demonstrated to be accurate in \cite{LuisMultiFR}. For simplicity in following expressions, we define $R_\mathcal{I} = \sum_{ i \in \mathcal{I} } R_i$ and $R_\mathcal{G} = \sum_{ g \in \mathcal{G} } R_g$.

The total synchronous inertia is the sum of inertia in each of the synchronous generators operating at a given time:
\begin{equation} \label{eq:SyncInertia}
    H_\textrm{sync} = \sum_{ g \in \mathcal{G} } \textrm{H}_g \cdot \textrm{P}^\textrm{max}_g \cdot y_g
\end{equation}

Synthetic inertia provided by wind turbines is based on extracting kinetic energy from the turbine, but this energy must be recovered a few seconds later so that the power output of the wind turbine can be equal to the pre-contingency value. Therefore, the `recovery effect' due to synthetic inertia is modelled as follows \cite{FeiAssessment}:
\begin{equation} \label{eq:RecoveryDefinition}
    \textrm{P}_\textrm{rec}(t)=\left\{ 
    \begin{array}{ll}
         0 \quad & \mbox{if $t \leq \textrm{T}_\textrm{rec}$} \\
        \textrm{k}_\textrm{rec} \cdot H_\textrm{synt} \quad & \mbox{if $ t > \textrm{T}_\textrm{rec} $}
    \end{array}
    \right.
\end{equation}
Where the value of `$\textrm{k}_\textrm{rec}$' is in fact a design parameter for the wind turbine controller, since the recovery will be higher if the turbine provides synthetic inertia for a longer period, and lower otherwise \cite{GE_AustraliaReport}.

GFM inverters can mimic the behaviour of synchronous machines, therefore these devices contribute to synthetic inertia:
\begin{equation} \label{eq:SyntInertia}
    H_\textrm{synt} = \sum_{i \in \textrm{GFM}} \textrm{H}_i \cdot \left( \textrm{P}_i - P^\textrm{curt}_i \right)
\end{equation}
Note that the synthetic inertia delivered to the system by a wind turbine can be reduced at the expense of curtailing some of its power output, through decision variable $P^\textrm{curt}_i$. 

On the other hand, GFL inverters have an inherent delay in increasing their power output due to a frequency drop in the system, as they must measure grid frequency through their Phase-Locked Loop before being able to react to a frequency change. Therefore, they cannot provide a power injection to the grid at the very instant of a generation loss (which is the exact instant when the frequency drop starts), so they are considered in this paper to contribute to the EFR service defined in eq.~(\ref{eq:EFRdefinition}), but unable to provide synthetic inertia.

\subsection{Constraints for respecting frequency limits} \label{sec:FreqConstraints}

To avoid disconnection of distributed generation, the maximum RoCoF must be limited below the threshold that would activate islanding protection devices, i.e.~$\textrm{RoCoF}_\textrm{max}$:
\begin{equation} \label{eq:Rocof}
    \textrm{RoCoF}_\textrm{max} =
    \left| \frac{\textrm{d} \Delta f(t=0)}{\textrm{d} t} \right|
    \geq
    \frac{ P_\textrm{L} \cdot f_0}{2\cdot (H_\textrm{sync} + H_\textrm{synt}) }
\end{equation}

To avoid the activation of Under-Frequency Load Shedding (UFLS), the frequency nadir must be contained above the triggering level of UFLS relays, i.e.~$\Delta f_\textrm{max}$:
\begin{equation} \label{eq:nadir}
    \left( \frac{ H_\textrm{sync} + H_\textrm{synt} } {f_0} - \frac{ R_\mathcal{I} \cdot \textrm{T}_\textrm{EFR} } {4 \Delta f_\textrm{max} } \right) 
    \frac{ R_\mathcal{G} }{ \textrm{T}_\textrm{PFR} } 
    \geq 
    \frac{ \left( P_\textrm{L} - R_\mathcal{I} \right) ^2 }{ 4 \Delta f_\textrm{max} }
\end{equation}
The deduction of the above constraint is included in Appendix~\ref{ap:NadirConstraint}. Note that the terms in constraint~(\ref{eq:nadir}) have been arranged to form a rotated Second-Order Cone (SOC), i.e. to take the structure `$x_1 \cdot x_2 \geq \left( x_3 \right)^2$', where $x_1,x_2 \geq 0$ \cite{MosekBook}.

Finally, a power equilibrium must be reached following the contingency, while considering the recovery effect for synthetic inertia:
\begin{equation} \label{eq:qss}
    R_\mathcal{I} + R_\mathcal{G} 
    \geq
    P_\textrm{L} + \textrm{k}_\textrm{rec} \cdot H_\textrm{synt}
\end{equation}
Eq.~(\ref{eq:qss}) is deduced from solving~(\ref{eq:SwingEq}) for `$t > \textrm{T}_\textrm{rec}$' and considering that frequency has stabilised after the contribution from inertia, EFR and PFR (i.e.~the frequency derivative is now zero).

Eqs.~(\ref{eq:Rocof}), (\ref{eq:nadir}) and (\ref{eq:qss}) are referred to in this paper as the RoCoF constraint, the nadir constraint, and the quasi-steady-state (q-s-s) constraint, respectively.

\subsection{Shadow prices for inertia and frequency response reserves}

The shadow prices for ancillary services for frequency control are obtained from the dual variables of the system constraints for respecting frequency limits, introduced in Section~\ref{sec:FreqConstraints}, following the procedure in \cite{LuisPricing}. The duals of the linear RoCoF and q-s-s constraints, eqs.~(\ref{eq:Rocof}) and (\ref{eq:qss}), are referred to as $\lambda_\textrm{RoCoF}$ and $\lambda_\textrm{q-s-s}$. The nadir constraint~(\ref{eq:nadir}) is a rotated SOC, therefore it is converted into standard SOC form before defining its dual variables:
\begin{multline} \label{eq:nadirSOC}
    \norm{
    \begin{bmatrix} 
        \frac{1}{f_0}  
        & \frac{-\textrm{T}_\textrm{EFR}}{4\Delta f_\textrm{max}}  
        &  \frac{-1}{\textrm{T}_\textrm{PFR}} 
        & 0 \\[5pt]
        0 &\frac{-1}{\sqrt{\Delta f_\textrm{max}}} & 0 &\frac{1}{\sqrt{\Delta f_\textrm{max}}}
    \end{bmatrix}
    \begin{bmatrix} 
        H_\textrm{sync} + H_\textrm{synt}   
        \\ R_\mathcal{I} 
        \\ R_\mathcal{G} 
        \\ P_\textrm{L} 
    \end{bmatrix} 
    } 
    \\
    \leq  \begin{bmatrix} 
        \frac{1}{f_0}  
        & \frac{-\textrm{T}_\textrm{EFR}}{4\Delta f_\textrm{max}}  
        & \frac{1}{\textrm{T}_\textrm{PFR}} 
        & 0 
    \end{bmatrix}
    \begin{bmatrix} 
        H_\textrm{sync} + H_\textrm{synt}   
        \\ R_\mathcal{I} 
        \\ R_\mathcal{G} 
        \\ P_\textrm{L} 
    \end{bmatrix} 
\end{multline}
The SOC dual variables of~(\ref{eq:nadirSOC}) are defined as $\lambda_1$ and $\lambda_2$ for the first and second rows in the left-hand side matrix of~(\ref{eq:nadirSOC}), and $\mu$ for the vector on the right-hand side. To comply with dual feasibility, the following constraint must be enforced on the dual variables:
\begin{equation}
    \norm{\begin{matrix}\lambda_1 \\ \lambda_2 \end{matrix}} \leq \mu
\end{equation}
The interested reader is advised to refer to \cite{ConvexOptPowerBook} for further details on dual formulations of SOCs. 

The Lagrangian function of any optimization problem including constraints~(\ref{eq:Rocof}), (\ref{eq:nadir}) and (\ref{eq:qss}) would then have the following terms, as explained in \cite{LuisPricing}:
\begin{multline} \label{eq:Lagrangian}
    \lambda_\textrm{RoCoF} \left[ \frac{P_\textrm{L} \cdot f_0}{\mbox{RoCoF}_\textrm{max}} - 2 \cdot \left( H_\textrm{sync} + H_\textrm{synt} \right) \right] \\
    +
    \lambda_\textrm{q-s-s} \left[ P_\textrm{L} + \textrm{k}_\textrm{rec} \cdot H_\textrm{synt} - (R_\mathcal{I} + R_\mathcal{G}) \right] \\
    +
    \lambda_1 \Biggl(
    \frac{ H_\textrm{sync} + H_\textrm{synt} }{f_0} 
    - \frac{ R_\mathcal{I} \cdot \textrm{T}_\textrm{EFR} } {4\Delta f_\textrm{max}}  
    - \frac{ R_\mathcal{G} } { \textrm{T}_\textrm{PFR} }  \Biggr) 
    + \lambda_2 \Biggl(
    \frac{ P_\textrm{L} - R_\mathcal{I} }{ \sqrt{\Delta f_\textrm{max}} }  
    \Biggr)
    \\
    - \mu \Biggl(
    \frac{H_\textrm{sync} + H_\textrm{synt}}{f_0} 
    - \frac{ R_\mathcal{I} \cdot \textrm{T}_\textrm{EFR} } {4\Delta f_\textrm{max}}  
    + \frac{ R_\mathcal{G} } { \textrm{T}_\textrm{PFR} }  \Biggr)
\end{multline}

Finally, the gradient of eq.~(\ref{eq:Lagrangian}) (i.e.~the Karush-Kuhn-Tucker stationarity condition) determines the shadow prices for the different ancillary services for frequency control:
\begin{itemize}
    \item Price for $H_\textrm{sync}$:
    \begin{equation} \label{eq:SyncH_price}
        \frac{\mu -\lambda_1}{f_0} + 2\lambda_\textrm{RoCoF}
    \end{equation}
    
    \item Price for $H_\textrm{synt}$:
    \begin{equation}
        \frac{\mu -\lambda_1}{f_0} 
        + 2\lambda_\textrm{RoCoF}
        - \lambda_\textrm{q-s-s} \cdot \textrm{k}_\textrm{rec} 
    \end{equation}
    
    \item Price for EFR (i.e.~$R_\mathcal{I}$): 
    \begin{equation} \label{eq:priceEFR}
        \frac{\left( \lambda_1 - \mu \right) \textrm{T}_\textrm{EFR}}{4\Delta f_\textrm{max}}
        + \frac{\lambda_2}{\sqrt{\Delta f_\textrm{max}}}
        + \lambda_\textrm{q-s-s}
    \end{equation}
    
    \item Price for PFR (i.e.~$R_\mathcal{G}$): 
    \begin{equation} \label{eq:PFR_price}
        \frac{\mu + \lambda_1}{\textrm{T}_\textrm{PFR}}
        + \lambda_\textrm{q-s-s}
    \end{equation}
    
\end{itemize}

The mathematical deductions in this Section~\ref{sec:Methodology} are general and would apply to any power system under consideration. While two simplifications are made here (i.e.~assuming that frequency is roughly equal in all of the network buses as in eq.~(\ref{eq:SwingEq}), and the linear approximation of droop control in eqs.~(\ref{eq:EFRdefinition}) and (\ref{eq:PFRdefinition})) they have been shown to be accurate in previous works \cite{OPFChavez,LuisMultiFR}. These two simplifications are necessary to obtain closed-form expressions for the frequency constraints deduced in Section~\ref{sec:FreqConstraints}. In turn, closed-form expressions of these constraints are necessary to compute shadow prices. Nonetheless, these simplifications are only needed for formulating the market clearing of ancillary services, while a full dynamic model would be used for tuning the real-time controls of the different devices providing frequency support to the grid. The frequency stability of the system would this way be fully guaranteed without the need for any simplification, while the shadow prices for ancillary services put in place incentives for investors to provide appropriate grid support, as demonstrated through relevant examples in Section~\ref{sec:CaseStudies}.

\section{Case studies} \label{sec:CaseStudies}

In this section, several case studies on a simplified Great Britain system are carried out to demonstrate the applicability of the proposed pricing formulation, and illustrate the incentives created for RES to provide ancillary services. This system is mostly based on some nuclear capacity, significant wind generation and a gas fleet.

For simplicity, the same level of demand is considered in all studies, of 25GWh (i.e.~25GW sustained for a period of 1h, which is the length assumed for the scheduling). The RES output is varied in the different case studies, to analyse its impact on the need for and value of ancillary services. A large nuclear plant of 1.8GW power rating is considered to be generating at all times, defining the size of the largest possible contingency (i.e.~the value of $P_\textrm{L}$). Technical characteristics of the different units in the generation mix are included in Table~\ref{tab:GenerationMix}. Note that the nuclear unit is assumed to provide no inertia, despite being a synchronous generator. This is due to the fact that the inertia from the nuclear generator would not be available in the event of its outage: given that this is the largest unit in the system driving the \textit{N}{-1} reliability requirement, its inertia cannot be relied upon when scheduling ancillary services.

The values for frequency limits used in these case studies correspond with the regulation in Great Britain, i.e. $\textrm{RoCoF}_\textrm{max}=1\textrm{Hz/s}$ and $\Delta f_\textrm{max}=0.8\textrm{Hz}$. The speed of the two frequency response services is defined as $\textrm{T}_\textrm{EFR}=1\textrm{s}$ and $\textrm{T}_\textrm{PFR}=10\textrm{s}$, while synthetic inertia is delivered until PFR has been fully delivered (i.e.~$\textrm{T}_\textrm{rec}$ is slightly greater than 10s). The optimisations were solved using the modelling layer YALMIP \cite{YALMIP} and Gurobi as the low-level numerical solver.

\begin{table}[t!]
    \captionsetup{justification=centering, textfont={sc,footnotesize}, labelfont=footnotesize, labelsep=newline}
    \centering
    \renewcommand{\arraystretch}{1.2}
    \caption{Characteristics of power generators}
    \label{tab:GenerationMix}
    \begin{tabular}{p{3.1cm}| p{0.8cm}| l| l}
                          & Nuclear     & Gas       & Wind \\ \hline
    Capacity installed (GW)       & 1.8           & 27.5        & 30 \\ \hline
    Power range per unit (MW)  ($\textrm{P}^\textrm{msg} -  \textrm{P}^\textrm{max}$)      & 1800-1800   & 250-550   & 0-2.5 \\ \hline
    No-load cost $\textrm{c}^\textrm{nl}_g$ (\pounds)      & 0           & 500       & 0 \\ \hline
    \multicolumn{1}{l|}{Marginal cost $\textrm{c}^\textrm{m}_g$ (\pounds/MWh)}  & 10          & 50        & 0 \\ \hline
    Inertia constant (s)  & N/A         & 5         & 0 or 5 \\ \hline
    Response type         & N/A         & PFR       & None or EFR \\ \hline
    Response capacity (MW)   & N/A         & $20\% \cdot \textrm{P}^\textrm{max}_g$       & 0 or $30\% \cdot \textrm{P}^\textrm{max}_i$ \\ \hline
    
    \end{tabular}
\end{table}

\begin{figure}[!t]
    \centering
    \includegraphics[width=0.81\columnwidth]{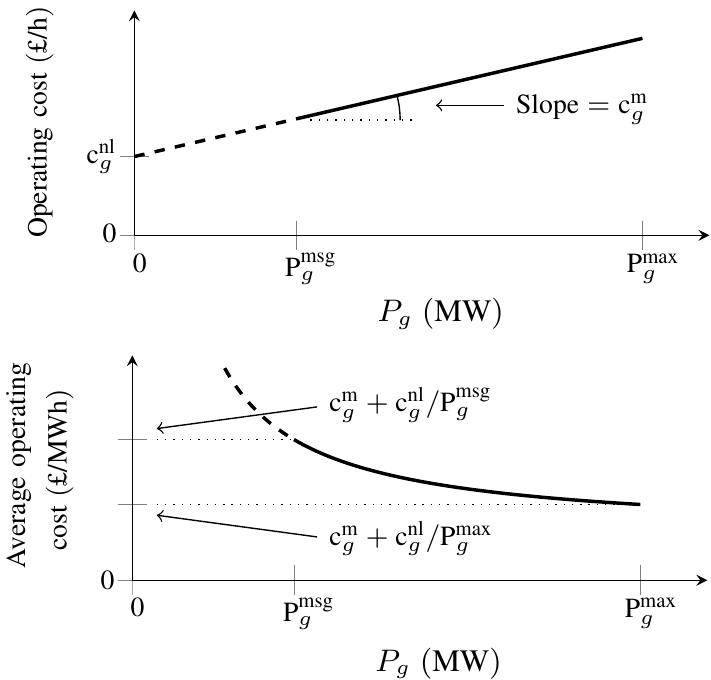}
    \vspace*{3mm}
    \caption{Graphical explanation of the operating cost and average operating cost of a thermal generator.} 
    \label{FigCostGenerators}
\end{figure}

No explicit cost for inertia or frequency response reserve is considered in these studies: the cost of these services is implicitly defined by the lower efficiency of part-loaded generators. To illustrate this point, a graphical description of the running costs of a thermal generator is provided in Fig.~\ref{FigCostGenerators}, assuming a linear cost curve. The average operating cost of a thermal unit always decreases with increasing loading level of the generator, being rated power $\textrm{P}_g^\textrm{max}$ the most efficient operating point. This implies that part-loading thermal units to provide frequency response reserve always increases total system costs, even if no explicit cost for this reserve is included; while the marginal generator could provide reserve at no additional opportunity cost, the total volume of system reserve required for containing frequency in the event of a large generator outage greatly exceeds the maximum headroom of any generator, as demonstrated in the case studies in following sections. 

The same reasoning applies to inertia: although inertia does not require headroom, given that a generator provides inertia simply by being synchronised to the grid, committing a higher number of part-loaded generators in order to increase system inertia increases the average fuel costs, when compared to the same amount of energy provided by a smaller number of fully-loaded generators. 

In a similar way, RES output that is curtailed to provide response implies a missed opportunity to produce more energy at zero fuel cost for the system. Regarding synthetic inertia, while this service does not require pre-contingency curtailment, its recovery effect translates into a higher requirement for frequency response reserves in eq.~(\ref{eq:qss}), which therefore also implies a cost for the system.

\subsection{Mathematical formulation of the Unit Commitment} \label{sec:UC_formulation}

The optimisation problem solved here is a frequency-secured Unit Commitment (UC), which simultaneously schedules energy production and ancillary services. For the sake of simplicity and clarity in showing the pricing scheme, a single-period UC is solved in each case, while the formulation is suitable to be extended to multi-period. 
\begin{alignat}{3}
    & {\text{min}} \quad
    && \sum_{g \in \mathcal{G}} \textrm{c}^\textrm{nl}_g\cdot y_g + \textrm{c}^\textrm{m}_g \cdot P_g  \label{ObjectiveFunction_UC}\\
    & \text{s.t.} \quad
    && \sum_{g \in \mathcal{G}}P_g + \sum_{\forall i} \left( \textrm{P}_i -P^\textrm{curt}_i \right) = \textrm{P}_\textrm{D} \label{LoadBalance_UC}\\
    & && y_g \in \{0, 1\} && \quad \forall g \in \mathcal{G} \label{Binary_y}\\
    & && y_g\cdot\textrm{P}^\textrm{msg}_g \leq P_g \leq y_g\cdot\textrm{P}^\textrm{max}_g && \quad \forall g \in \mathcal{G} \label{GenLimits_UC}\\
    & && 0 \leq R_g \leq y_g\cdot\textrm{R}^\textrm{max}_g && \quad \forall g \in \mathcal{G} \label{GenFR_Limits_UC}\\
    & && 0 \leq R_g \leq \textrm{P}^\textrm{max}_g-P_g && \quad \forall g \in \mathcal{G} \label{GenFR_Limits2_UC}\\
    & && 0 \leq P^\textrm{curt}_i \leq \textrm{P}_i  && \quad \forall i \in \mathcal{I} \label{REScurtailed_limit_UC}\\
    & && 0 \leq R_i \leq \textrm{R}^\textrm{max}_i  && \quad \forall i \in \mathcal{I}  \label{RES_FR_limit_UC}\\
    & && 0 \leq R_i \leq P^\textrm{curt}_i  && \quad \forall i \in \mathcal{I}  \label{RES_FR_limit2_UC}\\
    & && \textrm{Sync.} \; \textrm{inertia} \; \textrm{from} \; \textrm{all} \; g\textrm{,} \; \textrm{eq.} \; (\ref{eq:SyncInertia}) \nonumber\\
     & && \textrm{Synt.} \; \textrm{inertia} \; \textrm{from} \; \textrm{all} \; \textrm{GFM,} \; \; \textrm{eq.} \; (\ref{eq:SyntInertia}) \nonumber\\
    & && \textrm{RoCoF} \; \textrm{constraint,} \; \textrm{eq.} \; (\ref{eq:Rocof}) \nonumber\\
    & && \textrm{Nadir} \; \textrm{constraint,} \; \textrm{eq.} \; (\ref{eq:nadir}) \nonumber\\
    & && \textrm{q-s-s} \; \; \textrm{constraint,} \; \textrm{eq.} \; (\ref{eq:qss}) \nonumber 
\end{alignat}

The UC minimises fuel costs in the power system while meeting demand and ancillary services requirements. The load-balance constraint is defined in eq.~(\ref{LoadBalance_UC}). Thermal generators are modelled through the following constraints: eq.~(\ref{Binary_y}) enforces the integrality of their commitment state, eq.~(\ref{GenLimits_UC}) defines their power output limits, and eqs.~(\ref{GenFR_Limits_UC}) and (\ref{GenFR_Limits2_UC}) constrain the response provided from their headroom. RES are modelled with: eq.~(\ref{REScurtailed_limit_UC}), which limits the power output below the maximum available at any given time, and eqs.~(\ref{RES_FR_limit_UC}) and (\ref{RES_FR_limit2_UC}), which define the response capabilities of the RES units.

The optimisation problem defined above is a Mixed-Integer Second-Order Cone Program (MISOCP) due to the binary commitment decisions for thermal generators, i.e.~constraints (\ref{Binary_y}), and the SOC nadir constraint, while the rest of the constraints are linear. This type of problem can be efficiently solved even for large instances with off-the-shelf optimisation software for mixed-integer convex problems \cite{BoydConvex}. Note that in order to compute the prices of ancillary services as defined by eqs.~(\ref{eq:SyncH_price}) through (\ref{eq:PFR_price}), the dual form of the above optimisation problem must be solved, for which constraints~(\ref{Binary_y}) are relaxed (i.e.~substituted by `$y_g \in [0, 1]$'), therefore making the problem a convex SOC. Relaxing the commitment decision of thermal generators is one of the possibilities for computing electricity prices \cite{ConvexOptPowerBook}, since shadow prices can only be computed in a convex formulation, while here the original MISOCP formulation is solved to obtain the scheduling solution for the power system. Further discussion on alternative methods for computing prices under binary variables for the commitment of thermal generators is included in Section~\ref{subsection:sp_pricing}.

\subsection{Energy and ancillary services prices with no frequency support from RES} \label{sec:WindNoSupport}

\begin{figure}[!t]
    \centering
    \includegraphics[width=\columnwidth]{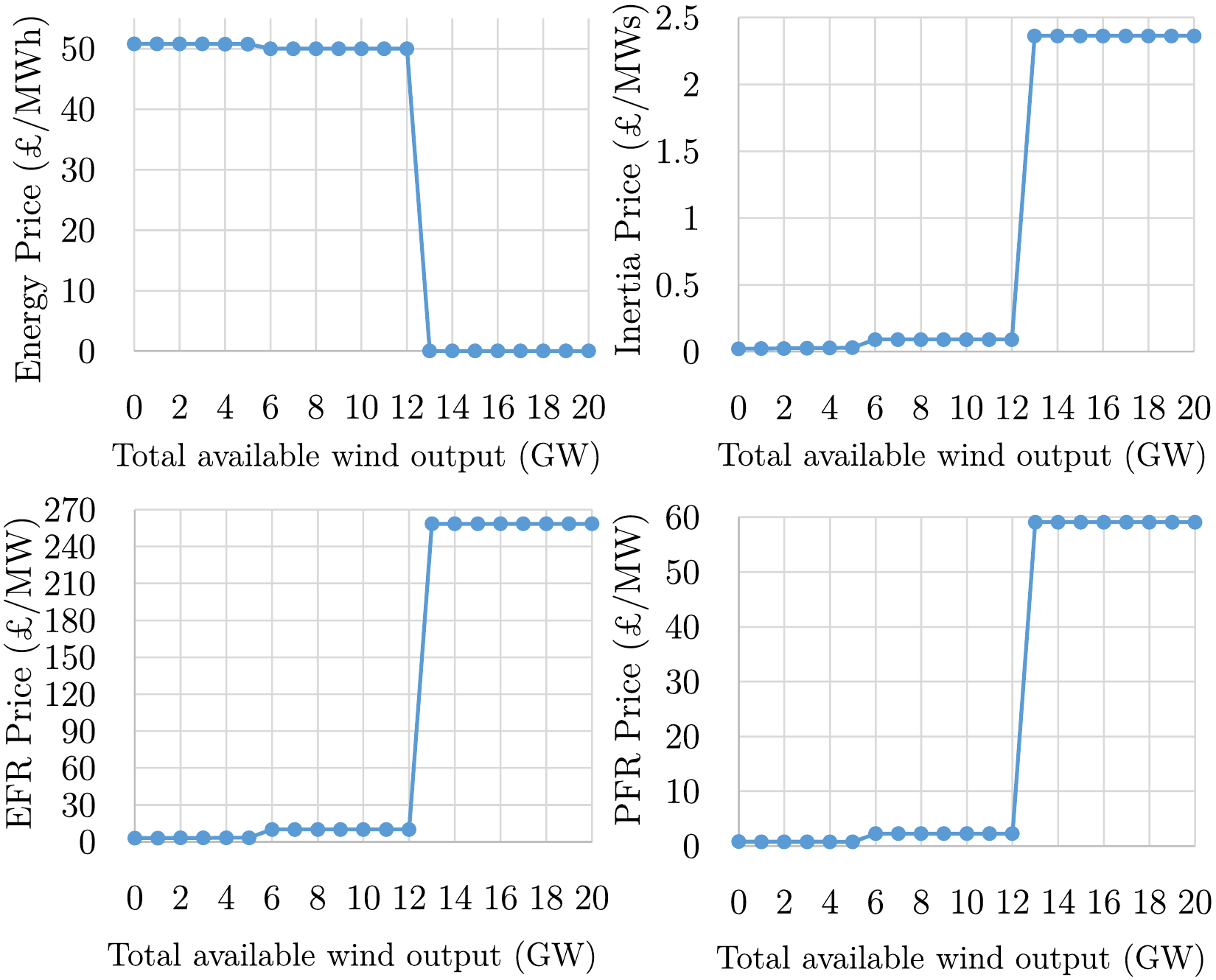}
    \caption{Prices of energy and ancillary services for the case of RES not contributing to provision of ancillary services.}
    \label{fig:WindNoSuport}
\end{figure}

First, we consider a case in which wind plants do not participate in provision of ancillary services, corresponding to current practice in most power grids in the world where non-synchronous RES are only used for energy production. Therefore, the inertia constant of wind is set to $\textrm{H}_i=0\textrm{s}$ and their frequency response capacity to $\textrm{R}^\textrm{max}_i=0\textrm{MW}$. The results are shown in Fig.~\ref{fig:WindNoSuport}, where the wind power output is progressively increased from 0GW up to 20GW, to analyse its effect on prices for energy and ancillary services. For a better understanding of the results, Tables~\ref{tab:RESnoSupport_0Wind} and \ref{tab:RESnoSupport_20Wind} present all details of the system operating condition for 0GW and 20GW of wind output, including costs, prices and revenues for the different plants.

\begin{table}[!t]
    \captionsetup{justification=centering, textfont={sc,footnotesize}, labelfont=footnotesize, labelsep=newline} 
    \renewcommand{\arraystretch}{1.2}
    \centering
    \caption{System operating condition for 0GW wind output}
    \label{tab:RESnoSupport_0Wind}
    \begin{tabular}{p{3.6cm}| l| l| l}
    
                              & Nuclear & Gas & Wind  \\ \hline
    Power produced (GW)       & 1.80    & 23.20      & 0          \\
    Units committed           & 1       & 50         & -          \\
    Response provided (GW)           & -       & 3.68       & 0          \\
    Operating cost (£k)       & 18.00   & 1,185.00    & 0          \\
    Revenue from energy (£k)  & 91.44   & 1,178.56    & 0          \\
    Revenue from response (£k)      & -       & 2.96       & 0          \\
    Revenue from inertia (£k) & -       & 3.12      & 0          \\ \hline
    Energy price (£/MWh)      & \multicolumn{3}{c}{50.80}        \\ \hline
    Inertia price (£/MWs)           & \multicolumn{3}{c}{0.02}         \\ \hline
    PFR price (£/MW)          & \multicolumn{3}{c}{0.80}         \\ \hline
    \end{tabular}
\end{table}

\begin{table}[!t]
    \captionsetup{justification=centering, textfont={sc,footnotesize}, labelfont=footnotesize, labelsep=newline} 
    \renewcommand{\arraystretch}{1.2}
    \centering
    \caption{System operating condition for 20GW wind output available}
    \label{tab:RESnoSupport_20Wind}
    \begin{tabular}{p{3.6cm}| l| l| l}
    
                              & Nuclear & Gas & Wind \\ \hline
    Power produced (GW)       & 1.80    & 10.25      & 12.95      \\
    Units committed           & 1       & 41         & -          \\
    Response provided (GW)           & -       & 4.49       & 0          \\
    Operating cost (£k)       & 18.00   & 533.00     & 0          \\
    Revenue from energy (£k)  & 0       & 0          & 0          \\
    Revenue from response (£k)      & -       & 265.31     & 0          \\
    Revenue from inertia (£k) & -       & 266.09     & 0          \\ \hline
    Energy price (£/MWh)      & \multicolumn{3}{c}{0}         \\ \hline
    Inertia price (£/MWs)           & \multicolumn{3}{c}{2.36}         \\ \hline
    PFR price (£/MW)          & \multicolumn{3}{c}{59.09}        \\ \hline
    \end{tabular}
\end{table}

Table~\ref{tab:RESnoSupport_0Wind} shows that all 50 gas generators are online if wind outputs 0GW, as these plants are needed to provide both energy, and inertia and PFR. The price of ancillary services is relatively low, only covering the opportunity costs of part-loaded gas plants that could provide more energy per plant if no PFR was needed. Regarding the impact of increasing wind output, Fig.~\ref{fig:WindNoSuport} demonstrates that all wind is curtailed above 13GW wind power: no more wind can be accommodated because at least 10.25GW of gas plants are needed to provide inertia and PFR to comply with frequency constraints. This situation is further detailed in Table~\ref{tab:RESnoSupport_20Wind}.

The price of energy, included in Fig.~\ref{fig:WindNoSuport}, is obtained from the dual variable of the load-balance constraint, eq.~(\ref{LoadBalance_UC}), and set by the operating cost of the most expensive generator needed for energy purposes at any given time: either gas plants or wind, once wind curtailment starts. The price of energy drops to \pounds0/MWh once wind curtailment is necessary, since the next MWh could be provided by wind with zero fuel costs. In this situation, synchronous generators would only be remunerated for ancillary services: they must produce energy when needed online for inertia and PFR purposes, due to their minimum stable generation (reason behind the wind curtailment), but their fuel costs would now be covered through their revenue from ancillary services, not from energy. Note that gas plants would need make-whole payments in Table~\ref{tab:RESnoSupport_20Wind}, a typical issue even in energy-only markets due to the lumped costs of thermal generators \cite{ConvexHullPricing}. Even though wind has zero revenue in Table~\ref{tab:RESnoSupport_20Wind} due to being partially curtailed, these units typically benefit from feed-in tariffs or similar financial instruments, receiving payments even if energy prices are zero. The nuclear unit would make a loss in this single-period UC, a realistic situation since nuclear plants typically have high startup costs and therefore can remain online during short periods of low energy prices to avoid turning on and off.

The price of ancillary services in Fig.~\ref{fig:WindNoSuport} increases very significantly once wind curtailment starts, since a higher volume of ancillary services available would reduce the need for turning on gas plants, therefore allowing more wind to be accommodated, with the corresponding savings in fuel costs for the system. The price of inertia increases by a factor 25, while the price of EFR is extremely high reaching \pounds260/MW. This price of EFR can be deduced from expression~(\ref{eq:priceEFR}), even though there is no EFR provider in this case. The very high price of this service in the wind-curtailment stage means that there is a high incentive for providers of EFR to enter the market. Wind plants with some curtailment, if appropriately controlled, could provide EFR and obtain revenue from this service, while a higher availability of EFR will have the effect of a decrease in price, as we show in the next section.

\subsection{Value of frequency response from curtailed RES} \label{sec:WindEFR}

\begin{figure}[!t]
    \centering
    \includegraphics[width=\columnwidth]{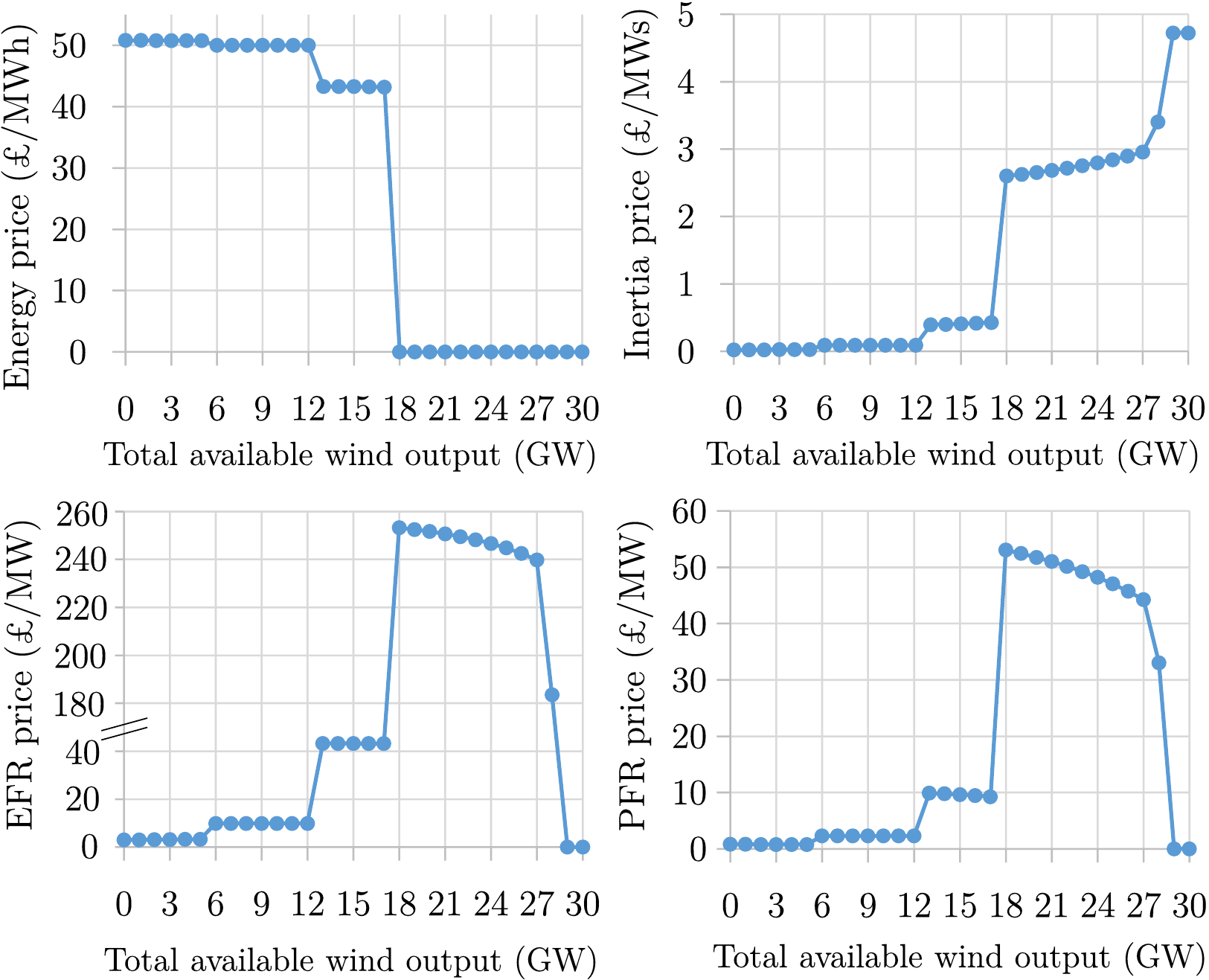}
    \caption{Prices of energy and ancillary services, with 15\% of the wind fleet having EFR capability.}
    \label{fig:WindOnlyEFR}
\end{figure}

Given the high incentives for EFR shown in the previous section, here we consider that 15\% of the wind fleet has the capability to provide EFR when part-loaded. Wind turbines equipped with GFL inverters are sufficient for this purpose, as long as the control of the inverters includes a droop characteristic, which increases the power output of the turbine in the event of a frequency drop in the grid.      

Compared to the case shown in Section~\ref{sec:WindNoSupport}, the results in Fig.~\ref{fig:WindOnlyEFR} demonstrate that an additional 5GW of wind can be accommodated, since the EFR provided by 15\% of the wind fleet implies that a lower number of gas plants are needed online to provide inertia and PFR. Table~\ref{tab:WindEFR_20Wind} shows that some wind is curtailed to provide this EFR, as headroom is needed for this service, but overall wind curtailment is reduced due to the fact that any wind curtailed in  Section~\ref{sec:WindNoSupport} did not contribute to supporting frequency. Furthermore, the portion of the wind fleet capable of providing EFR would potentially capture a very significant revenue, as illustrated in Table~\ref{tab:WindEFR_20Wind}, while energy-only wind would need out-of-market support due to the \pounds0/MWh energy price.

\begin{table}[!t]
    \captionsetup{justification=centering, textfont={sc,footnotesize}, labelfont=footnotesize, labelsep=newline} 
    \renewcommand{\arraystretch}{1.2}
    \centering
    \caption{System operating condition for 20GW wind output available, 15\% of which has EFR capability}
    \label{tab:WindEFR_20Wind}
    \begin{tabular}{p{3.2cm}| l| l| p{0.9cm}| p{0.8cm}}
    
                              & Nuclear & Gas & Wind no-EFR & Wind EFR \\ \hline
    Power output (GW)           & 1.80    & 6.00       & 15.89       & 1.31        \\
    Units committed             & 1       & 24         & -           & -           \\
    Response provided (GW)             & -       & 2.43       & 0           & 0.90        \\
    Operating cost (£k)         & 18.00   & 312.00     & 0           & 0           \\
    Revenue from energy (£k)    & 0       & 0          & 0           & 0           \\
    Revenue from response (£k)        & -       & 125.78      & 0           & 226.49      \\ 
    Revenue from inertia (£k)   & -       & 175.56     & 0           & 0           \\ \hline
    Energy price (£/MWh)        & \multicolumn{4}{c}{0}                        \\ \hline
    Inertia price (£/MWs) & \multicolumn{4}{c}{2.66}                        \\ \hline
    EFR price (£/MW)            & \multicolumn{4}{c}{251.66}                      \\ \hline
    PFR price (£/MW)            & \multicolumn{4}{c}{51.76}                       \\ \hline
    \end{tabular}
\end{table}

Fig.~\ref{fig:WindOnlyEFR} shows several distinct levels for the price of EFR. For the range of 0GW to 12GW of wind power output, the price of EFR is fairly low, reaching a maximum of £9.9/MW. In fact, no EFR is available in these cases, since it is optimal to use all wind for energy purposes, due to the high level of inertia provided by synchronous generators that are also needed for energy. The second range of interest corresponds to 13GW to 17GW of wind output: in this range, it is optimal to start curtailing some of the wind with EFR capability to provide this service, due to the decreasing inertia. The price of EFR is of £43/MW in this range, the same as the price of energy, since one extra MW of EFR would reduce wind curtailment by 1MW, therefore reducing the operating cost of the system by the exact amount that it costs to produce 1MWh. The third level for EFR prices is seen for wind availability of between 18GW and 27GW: in this range, all the EFR capability of curtailed wind is used (i.e. $30\% \cdot \textrm{P}^\textrm{max}_i$ as defined in Table~\ref{tab:GenerationMix}). The price of EFR reaches a highest value of £253/MW, and progressively decreases as more EFR becomes available with increasing wind power available. Then, there is a significant drop in price of EFR for a wind availability of 28GW: at this point the RoCoF constraint also becomes binding, requiring a higher volume of inertia that in turn lowers the need for EFR to comply with the nadir constraint, therefore decreasing its price. Finally, for wind availability above 28GW, the nadir constraint becomes inactive due to an excess of EFR, making the price of this service zero.

For the cases with highest wind in Fig.~\ref{fig:WindOnlyEFR}, the price of inertia reaches a very high value of £4.7/MWs. This high price for inertia incentivises the introduction of GFM inverters for synthetic inertia, which is investigated in the next section.

\subsection{Value of synthetic inertia from GFM inverters}
\label{subsection:val_SI_GFM}

The high prices for inertia seen in Section~\ref{sec:WindEFR} would promote synthetic inertia providers to enter the market. For isolating the effect of synthetic inertia on system operation and prices, here we consider that 30\% of the wind fleet is equipped with GFM inverters and therefore has the capability to provide synthetic inertia, while the remaining 70\% of wind only provides energy. A synthetic inertia constant of $\textrm{H}_i=5\textrm{s}$ is used in this section.

\begin{table}[!t]
    \captionsetup{justification=centering, textfont={sc,footnotesize}, labelfont=footnotesize, labelsep=newline} 
    \renewcommand{\arraystretch}{1.2}
    \centering
    \caption{System operating condition for 20GW wind output available, 30\% of which has synthetic inertia capability}
    \label{tab:WindGFM_20Wind}
    \begin{tabular}{p{3.2cm}| l| l| p{1cm}| p{0.7cm}}
                                & Nuclear & Gas & Wind no-GFM & Wind GFM \\ \hline
    Power output (GW)           & 1.80    & 9.00       & 8.20        & 6.00        \\ 
    Units committed             & 1       & 36         & -           & -           \\ 
    Response provided (GW)             & -       & 3.92       & 0           & 0           \\ 
    Operating cost (£k)         & 18.00      & 468.00     & 0           & 0           \\ 
    Revenue from energy (£k)    & 0       & 0          & 0           & 0           \\ 
    Revenue from response (£k)        & -       & 262.29     & 0           & 0           \\ 
    Revenue from inertia (£k)   & -       & 202.95     & 0           & 61.50       \\ \hline
    Energy price (£/MWh)        & \multicolumn{4}{c}{0}                        \\ \hline
    Sync inertia price (£/MWs) & \multicolumn{4}{c}{2.05}                        \\ \hline
    Synt inertia price (£/MWs)   & \multicolumn{4}{c}{2.05}                        \\ \hline
    EFR price (£/MW)            & \multicolumn{4}{c}{260.81}                      \\ \hline
    PFR price (£/MW)            & \multicolumn{4}{c}{66.91}                       \\ \hline
    \end{tabular}
\end{table}

\begin{figure}[!t]
    \centering
    \includegraphics[width=\columnwidth]{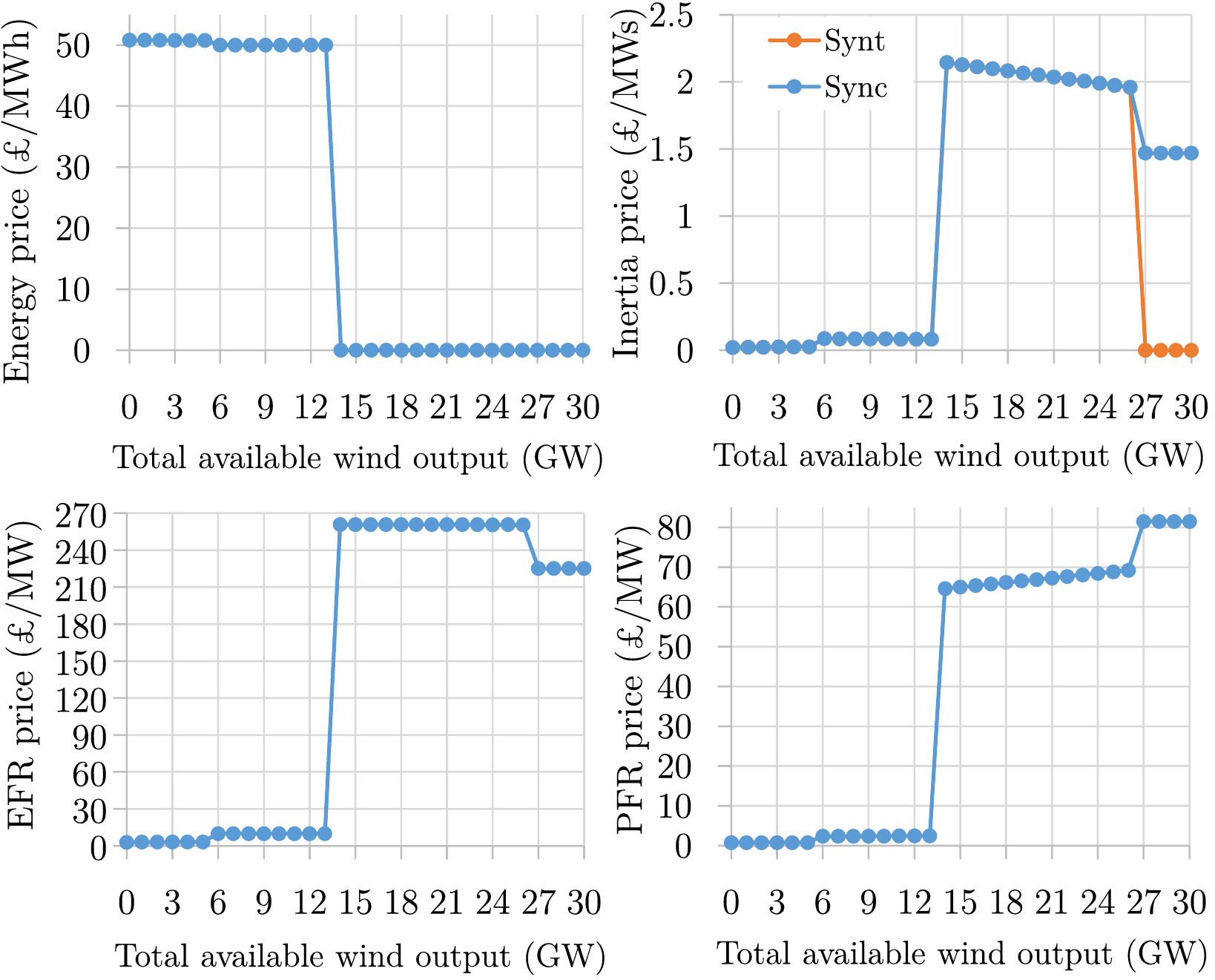}
    \caption{Prices of energy and ancillary services, with 30\% of the wind fleet having synthetic inertia capability and $\textrm{k}_\textrm{rec}=0.05\textrm{s}^{-1}$.}
    \label{fig:WindGFM}
\end{figure}

The results in Fig.~\ref{fig:WindGFM} show a decreasing trend for the price of inertia under high values of wind output, as compared to the flat prices seen in Fig.~\ref{fig:WindNoSuport}, where 100\% of the wind fleet was energy-only. This drop in inertia prices is due to the higher liquidity for this service as synthetic inertia providers enter the market. Nonetheless, GFM-equipped wind generators would make an important profit, as shown in Table~\ref{tab:WindGFM_20Wind}. Investment in GFM inverters would then be justified by the revenue stream created by this proposed pricing scheme for ancillary services from RES.

\begin{figure}[!t]
    \centering
    \includegraphics[width=\columnwidth]{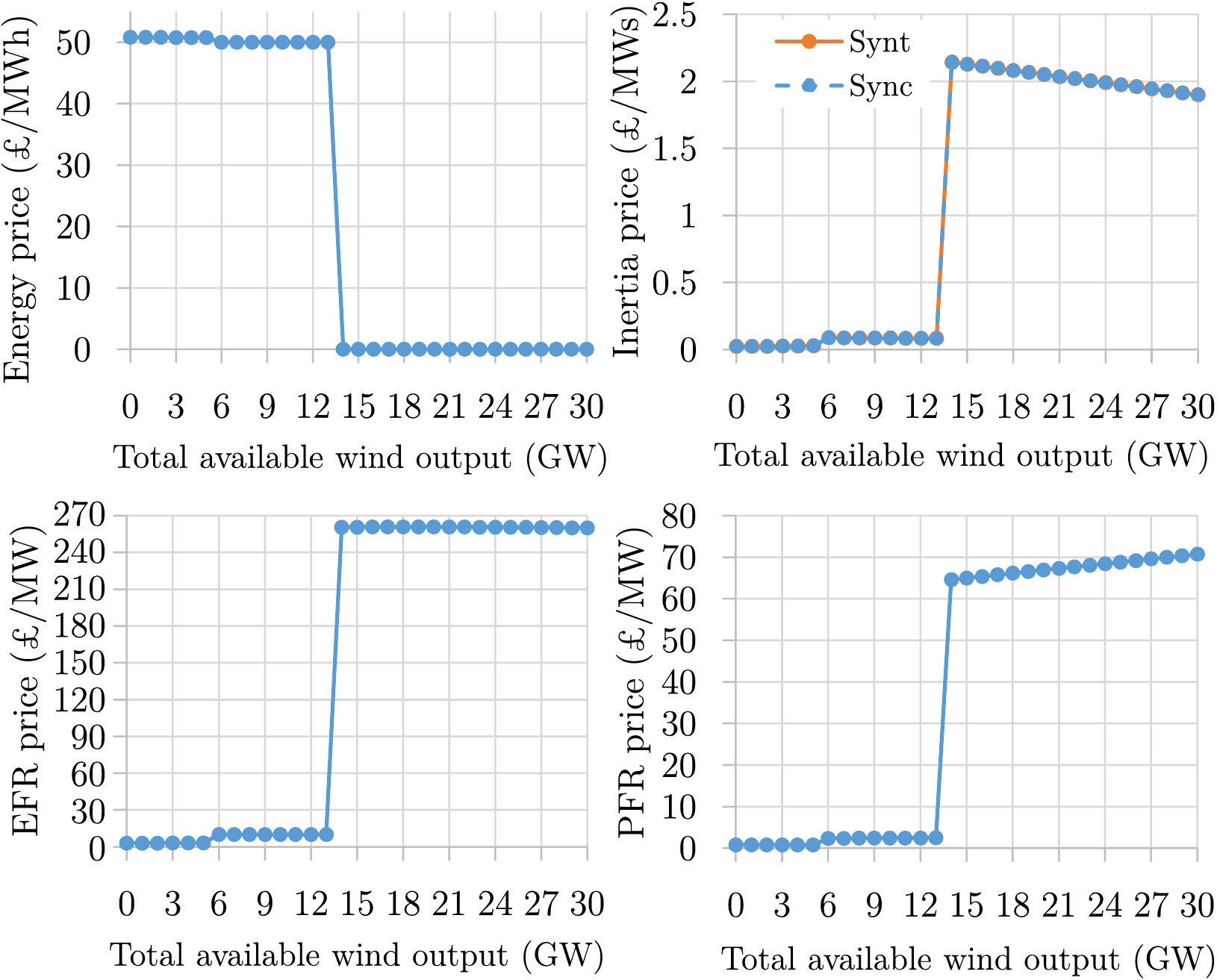}
    \caption{Prices of energy and ancillary services, with 30\% of the wind fleet having synthetic inertia capability and $\textrm{k}_\textrm{rec}=0.025\textrm{s}^{-1}$.}
    \label{fig:WindGFM_reco0025}
\end{figure}

Fig.~\ref{fig:WindGFM} also shows that the price of synthetic inertia drops to zero for wind output above 27GW, while the price of synchronous inertia is still positive. The reason for this difference is the recovery effect of synthetic inertia: once the penetration of synthetic inertia in the system is sufficiently high, the necessary frequency response reserves to compensate for the recovery effect in the quasi-steady-state constraint, eq.~(\ref{eq:qss}), becomes too high. After this point, adding more synthetic inertia does not lead to a decrease in system operating costs, hence the price of \pounds0/MWs for this service. Note that a RES unit with additional energy storage, such as solar PV panels with an ultracapacitor bank \cite{EasyRES}, could provide synthetic inertia without a recovery effect, and therefore would capture the `synchronous inertia' price. Battery storage systems with GFM inverters would also capture this price.

On the other hand, we analyse in Fig.~\ref{fig:WindGFM_reco0025} a case with a smaller recovery effect for the turbines, which could be achieved by either designing turbines with a higher physical inertia, or through enhanced wind-turbine controls to extract only the kinetic energy needed to secure the frequency nadir \cite{Birmingham_nadir}. In this case, the price of synthetic inertia does not drop to zero.

Since the prices of EFR again reach very high values in the case presented in this section, due to the lack of wind turbines providing this service, it is clear that a mix between GFM-equipped turbines for synthetic inertia and GFL-equipped turbines with an EFR control loop is necessary. Furthermore, a combination between synthetic inertia and EFR from RES would be needed for achieving 100\% inverter-based penetration from a frequency-stability perspective, as otherwise thermal plants would be needed to compensate the recovery effect via PFR, or to provide synchronous inertia for securing the maximum RoCoF.

\subsection{Synergy between synthetic inertia and frequency response from RES}

\begin{table}[!t]
    \captionsetup{justification=centering, textfont={sc,footnotesize}, labelfont=footnotesize, labelsep=newline} 
    \renewcommand{\arraystretch}{1.2}
    \centering
    \caption{System operating condition for 30GW wind output available, 60\% of which has EFR capability and 30\% synthetic inertia capability}
    \label{tab:FinalCaseStudy1}
    \begin{tabular}{p{3.2cm}| p{0.78cm}| p{0.35cm}| p{0.55cm}| p{0.62cm}| p{0.62cm}}
        & Nuclear & Gas & Wind & Wind EFR & Wind GFM \\ \hline
    Power output (GW)   & 1.80  & 0 & 3.00  & 11.20 & 9.00   \\
    Units committed & 1 & 0 & - & - & - \\ 
    Response provided (GW)  & - & 0 & 0 & 4.05  & 0 \\ 
    Operating cost (£k) & 18.00 & 0 & 0 & 0 & 0 \\ 
    Revenue from energy (£k)    & 0 & 0 & 0 & 0 & 0 \\ 
    Revenue from response (£k)  & - & 0 & 0 & 0    & 0 \\
    Revenue from inertia (£k)   & - & 0 & 0 & 0 & 212.85    \\ \hline
    Energy price (£/MWh)        & \multicolumn{5}{c}{0} \\ \hline
    Sync inertia price (£/MWs) & \multicolumn{5}{c}{4.73} \\ \hline
    Synt inertia price (£/MWs)   & \multicolumn{5}{c}{4.73} \\ \hline
    EFR price (£/MW)            & \multicolumn{5}{c}{0}  \\ \hline
    PFR price (£/MW)            & \multicolumn{5}{c}{0}  \\ \hline
    \end{tabular}
\end{table}

\begin{table}[!t]
    \captionsetup{justification=centering, textfont={sc,footnotesize}, labelfont=footnotesize, labelsep=newline} 
    \renewcommand{\arraystretch}{1.2}
    \centering
    \caption{System operating condition for 30GW wind output available, 40\% of which has synthetic inertia capability and 46\% EFR capability}
    \label{tab:FinalCaseStudy2}
    \begin{tabular}{p{3.2cm}| p{0.78cm}| p{0.35cm}| p{0.55cm}| p{0.62cm}| p{0.62cm}}
        & Nuclear & Gas & Wind & Wind EFR & Wind GFM \\ \hline
    Power output (GW)   & 1.80  & 0 & 4.22  & 9.65 & 9.33   \\
    Units committed & 1 & 0 & - & - & - \\ 
    Response provided (GW)  & - & 0 & 0 & 4.13  & 0 \\ 
    Operating cost (£k) & 18.00 & 0 & 0 & 0 & 0 \\ 
    Revenue from energy (£k)    & 0 & 0 & 0 & 0 & 0 \\ 
    Revenue from response (£k)  & - & 0 & 0 & 311.24    & 0 \\
    Revenue from inertia (£k)   & - & 0 & 0 & 0 & 0    \\ \hline
    Energy price (£/MWh)        & \multicolumn{5}{c}{0} \\ \hline
    Sync inertia price (£/MWs) & \multicolumn{5}{c}{0} \\ \hline
    Synt inertia price (£/MWs)   & \multicolumn{5}{c}{0} \\ \hline
    EFR price (£/MW)            & \multicolumn{5}{c}{75.36}  \\ \hline
    PFR price (£/MW)            & \multicolumn{5}{c}{75.36}  \\ \hline
    \end{tabular}
\end{table}

Given that both synthetic inertia and frequency response reserve have been shown to be valuable services during conditions of high RES output, here we investigate the combinations of GFM and GFL working in tandem to achieve covering all the demand with RES, from a frequency stability perspective. In Table~\ref{tab:FinalCaseStudy1}, a system condition with 30GW of wind output is considered, of which 10\% provides energy but no frequency-control services, 30\% provides synthetic inertia, and the remaining 60\% has the capability to provide EFR. The market outcome shows the highest price seen so far in these case studies for synthetic inertia, of £4.73/MWs, while the price for frequency response reserves is zero due to an excess of EFR. On the other hand, the case in Table~\ref{tab:FinalCaseStudy2} shows a system condition with again 30GW of wind output, but 40\% providing synthetic inertia, 46\% providing EFR, and the rest providing only energy. In this case, the over-supplied service is synthetic inertia, whose price drops to zero, while the price of EFR reaches a substantial value of more than £75/MW.

It is relevant to mention that in both cases included in Tables~\ref{tab:FinalCaseStudy1} and \ref{tab:FinalCaseStudy2}, the nadir constraint is never binding: prices of inertia and frequency response reserves are determined from the dual variables of the RoCoF and q-s-s constraints. The reason for nadir never being binding is that, to guarantee a 100\% RES penetration from a frequency-stability perspective, inertia must be purely synthetic, which implies a non-zero recovery effect in the q-s-s constraint. At the same time, the q-s-s constraint must be fully met by EFR coming from wind (no PFR coming from thermal plants can be used). This implies that the volume of EFR available must be greater than the size of the largest loss $P_\textrm{L}$, since EFR also has to compensate the recovery effect of synthetic inertia in the q-s-s constraint. Such a high volume of EFR, in combination with the synthetic inertia required to meet the RoCoF constraint, makes the nadir constraint inactive in these cases.

A future with frequent occurrences of 100\% RES penetration could see prices moving in between the values on Tables~\ref{tab:FinalCaseStudy1} and \ref{tab:FinalCaseStudy2}, that is, experiencing either high prices for synthetic inertia or high prices for frequency response reserve, depending on the availability of each service. For example, if wind farms equipped with GFM converters are mostly clustered in a certain region of the network where wind availability happens to be low for a given hour, while wind gusts are strong in another part of the network where wind turbines have the capability to provide EFR, a market outcome as the one shown in Table~\ref{tab:FinalCaseStudy1} could be realised. In this highly-renewable future, it is also likely that some periods with zero prices for all frequency-control services would be seen, due to an excess of both synthetic inertia and frequency response reserve. In this context, the proposed pricing scheme would allow the market to find an equilibrium on how much GFM capacity and how much GFL capacity with the capability to provide frequency response is needed.

Finally, it is relevant to point out that the prices shown in this section are based on the swing-equation model for frequency dynamics described in Section~\ref{sec:SwingModel}. While grids with 100\% penetration of inverter-based resources could show very different dynamics to the swing equation, there is some consensus in the literature that grid-forming inverters should be controlled to approximately mimic the inertial behaviour of synchronous machines, in which case the swing equation will remain an appropriate mathematical model for frequency dynamics \cite{XavierGuillaudSwing}. However, it is certainly possible that some future inverter-dominated grids are actually controlled in a different way which does not replicate the swing equation. If this is the case, the proposed methodology for pricing ancillary services would not directly apply during times of 100\% RES penetration, and further work would be needed to assign appropriate prices during these hours. This new methodology to price ancillary services should be based on the mathematical description of the system frequency dynamics, which will depend on the control strategy chosen for the inverters.

\subsection{Value and implications of optimising synthetic inertia provision under wind forecast error}

\begin{figure}[!t]
    \includegraphics[width=\columnwidth]{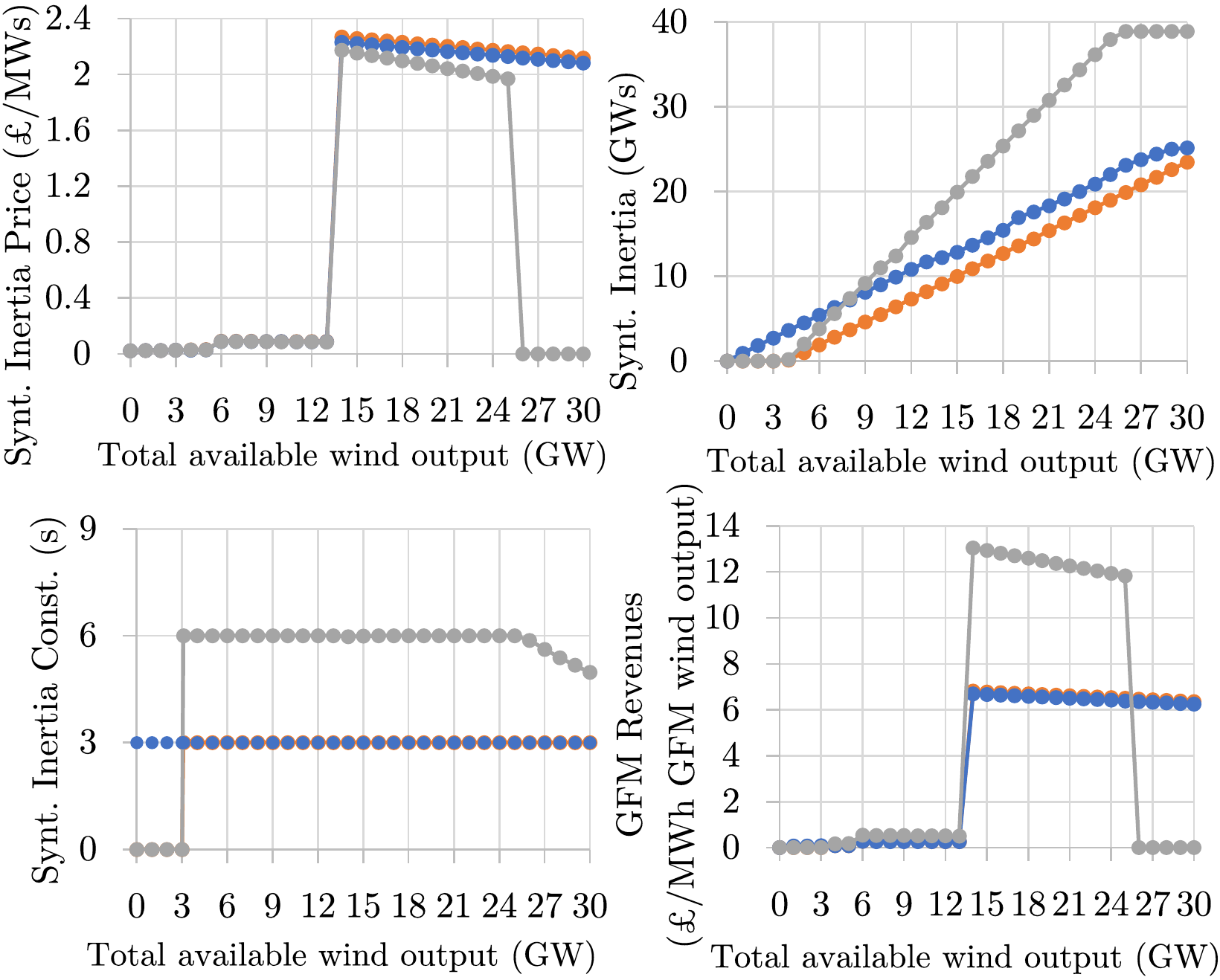}
    \caption{Implications of accounting for wind forecast error and benefits of optimising the synthetic inertia constant, $H_i$. Three cases are shown: 1) Forecast error neglected, with $\textrm{H}_i=3\textrm{s}$ (\protect\includegraphics[height=0.55em]{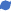}); 2) Forecast error of $\alpha = 13\%$, with $\textrm{H}_i=3\textrm{s}$ (\protect\includegraphics[height=0.55em]{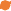}); 3) Forecast error of $\alpha = 13\%$, with $H_i$ optimised (\protect\includegraphics[height=0.55em]{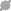}). }
    \label{fig:ForecastError_OptSyntInertia}
\end{figure}

We now analyse the impact of wind forecast error on the ability of this RES technology to provide synthetic inertia, as this is currently one of the main barriers for RES generators to participate in ancillary services markets. Day-ahead wind forecast errors on a national level have been reported to be of a maximum of 13.5\% of installed capacity for the Nordic countries \cite{ForecastErrorNordics}, and an average of between 9 to 12\% of installed capacity for continental Europe \cite{ForecastErrorENTSOE}. We therefore use a value of `$\alpha=13\%$' for the day-ahead wind forecast error, and show results for day-ahead prices for ancillary services considering that wind providers must be conservative in their volume offered to the market, given that their actual availability can potentially be lower than forecast.

To account for the wind forecast error, eq.~(\ref{eq:SyntInertia}) in the optimisation problem becomes: 
\begin{equation} 
    H_\textrm{synt} = \sum_{i \in \textrm{GFM}} \textrm{H}_i \cdot \left( \textrm{P}_i -P^\textrm{curt}_i - \alpha \cdot \textrm{P}_i^\textrm{max} \right)
\end{equation}

The results from a simulation with 30\% of the wind fleet equipped with GFM inverters and providing synthetic inertia with a constant of `$\textrm{H}_i=3\textrm{s}$' are shown in Fig.~\ref{fig:ForecastError_OptSyntInertia}. Due to considering the forecast error, a lower volume of synthetic inertia is available, which results in slightly higher prices for this service. 
Given this higher incentive for synthetic inertia providers, it would become relevant to optimise the synthetic inertia constant as a function of RES output in the system, in order to potentially provide more synthetic inertia when this service is very valuable. While in previous sections a fixed value for $\textrm{H}_i$ has been considered, this is in fact a design parameter, as the inertia constant of grid-forming inverters is defined via software and therefore can be tuned \cite{XavierGuillaudSwing}. The advantage of frequently updating the value of the synthetic inertia constant is to be able to reduce the recovery effect at times, or instead increase the synthetic inertia provision if a bigger recovery effect can be accommodated by the system cost-effectively. 

To optimise the volume of synthetic inertia delivered to the system, one can reformulate eq.~(\ref{eq:SyntInertia}) in the Unit Commitment to turn `$H_i$' into a decision variable:
\begin{equation} 
    H_\textrm{synt} = \sum_{i \in \textrm{GFM}} H_i \cdot \left( \textrm{P}_i - \alpha \cdot \textrm{P}_i^\textrm{max} \right)
\end{equation}
Where $H_i$ is now a continuous decision variable with bounds `$0\textrm{s} \leq H_i \leq 6\textrm{s}$' in the case studies shown here. To avoid a non-convexity in the above equality constraint, the curtailment of wind power from GFM units has been removed as a decision variable. Note that curtailment could still be done in energy-only wind generators, as defined by constraint~(\ref{REScurtailed_limit_UC}).

Fig.~\ref{fig:ForecastError_OptSyntInertia} also displays the results for the case where the value of the synthetic inertia constant is optimised. The optimal value for this variable is of $H_i=6\textrm{s}$ 
for levels of up to 26GW of wind power available (after this point, $H_i$ begins to decline as the higher recovery effect overtakes the value of additional synthetic inertia).
Given the higher available volume of this service compared to the cases where the value is fixed to $\textrm{H}_i=3\textrm{s}$, the price of synthetic inertia decreases; however, GFM-equipped wind generators that optimise their inertia constant capture a significantly higher revenue for this service, roughly twice as much as GFM assets with a fixed constant of 3s, for wind output up to 26GW.

It is important to note that this modelling exercise considers that $H_i$ can be optimised by the same entity that solves the central dispatch. Therefore, for wind power available above 26GW, the price of synthetic inertia in the `optimised' case drops to zero, as a higher volume of this service is no longer cost-effective for the system due to the recovery effect. This in turn causes the revenues of GFM units with an optimised inertia constant to drop to zero. However, this situation would change if this variable were to be privately optimised by GFM unit owners, which could even lead to strategic market behaviour. The problem of profit maximisation with strategic behaviour of some market participants is however beyond the scope of this work.

Finally, it is worth remarking that updating the synthetic inertia constant on an hourly or even sub-hourly basis 
implies that a communication system must be in place between the operator of a wind farm and the controller of each of the turbines. While deploying such communication network certainly implies some investment cost, the higher revenues perceived by optimising the value of the synthetic inertia delivered to the system, as shown in Fig.~\ref{fig:ForecastError_OptSyntInertia}, could create a sensible business case for this approach.

\subsection{Discussion on different pricing methodologies for dealing with non-convexities}
\label{subsection:sp_pricing}

It is relevant to discuss the compatibility of the shadow prices for frequency-control services from RES proposed in the present paper with the different pricing methodologies previously proposed for dealing with non-convexities in electricity markets. In this section we discuss three of these methodologies, namely the `dispatchable', `restricted' and `convex hull' pricing frameworks.

Dispatchable pricing is the methodology used in all previous case studies, which implies relaxing constraints~(\ref{Binary_y}) to `$y_g \in [0, 1]$'. This approach has the drawback of not fully representing the technical characteristics of thermal generators, therefore prices computed through this method might not correspond to a feasible operating point for the power system. 

Restricted pricing is computed as follows. First, the original mixed-integer problem described in Section~\ref{sec:UC_formulation} is solved. Then, the commitment decision from the solution of the mixed-integer problem, i.e. the optimal values for variables `$y_g$', are stored, referred to as `$y_g^*$'. The optimisation problem is solved for a second time, this time replacing constraints~(\ref{Binary_y}) by constraints `$y_g=y_g^*$'.  

In the restricted pricing method, a new term is added to the Lagrangian function~(\ref{eq:Lagrangian}): `$\lambda_\textrm{commit} \cdot (y_g^*-y_g)$'. The dual variable $\lambda_\textrm{commit}$ represents a `price for commitment', which creates a side payment for units associated to a commitment decision variable. The main advantage of this methodology is that it models the actual operation of the power system when computing the optimal value for dual variables.

Convex hull pricing was conceived to minimise uplift payments \cite{ConvexHullPricing}. These are side payments given to generators that fail to recover their costs through energy revenues, in order to ensure that all market participants follow the system operator's dispatch. The need for uplift payments arises from the fact that the Unit Commitment problem contains non-convexities, due to the aforementioned binary commitment variables for thermal generators. These non-convexities appear in the so-called `private constraints', which include the no-load costs defined through binary variables as in eq.~(\ref{Binary_y}). Convex hull prices (CHP) can be determined using a Lagrangian relaxation of the `system-wide constraints', which in the UC solved in this paper include the load-balance, RoCoF, nadir and q-s-s constraints, defined in eqs. (\ref{LoadBalance_UC}), (\ref{eq:Rocof}), (\ref{eq:nadir}) and (\ref{eq:qss}), respectively. As the RoCoF and q-s-s constraints are linear, they do not introduce any non-convexities. The nadir constraint is a Second-Order Cone, therefore it does not introduce any non-convexities either \cite{andersen2002notes}.

Now we demonstrate that CHP for the UC considered in this paper are equivalent to the prices obtained through the dispatchable method. First, as in energy-only UC formulations, the only source of non-convexities in this frequency-secured UC framework remains within the set of private constraints for thermal generators, `$\mathcal{X}_g$':
\begin{equation} \label{eq:feas_set}
    \mathcal{X}_g = \{P_g \in \mathbb{R}, R_g \in \mathbb{R}, y_g \in \{0,1\} \vert (\ref{GenLimits_UC}), (\ref{GenFR_Limits_UC}), (\ref{GenFR_Limits2_UC}) \}
\end{equation}

Then, as stated in \cite{HuaCHP}, by relaxing the binary commitment constraint to $y_g \in \mathbb{R}$ and bounding $y_g$ by its integer values, the convex hull for all convex combinations of the set of thermal generators $g$ is given by `$\textrm{conv}(\mathcal{X}_g)$':
\begin{multline} \label{eq:ch_set}
    \textrm{conv}(\mathcal{X}_g) = \\
    \{P_g \in \mathbb{R}, R_g \in \mathbb{R}, y_g \in \mathbb{R} \vert 0 \leq y_g \leq 1, (\ref{GenLimits_UC}), (\ref{GenFR_Limits_UC}), (\ref{GenFR_Limits2_UC}) \}
\end{multline}

We use here the CHP-primal methodology presented in \cite{HuaCHP}, where the individual cost function of each unit $g$ is the following linear function taken over $\mathcal{X}_g$:
\begin{equation} \label{eq:cost_func}
    C_g(P_g, R_g, y_g) = \textrm{c}^\textrm{nl}_g\cdot y_g + \textrm{c}^\textrm{m}_g \cdot P_g
\end{equation}
In this CHP-primal approach, each cost function $C_g(\cdot)$ is replaced by its convex envelope taken over the non-convex feasible set $\mathcal{X}_g$, i.e.~the largest convex function on `$\textrm{conv}(\mathcal{X}_g)$' that underestimates $C_g(\cdot)$ on $\mathcal{X}_g$. Given that we consider generators with a constant marginal cost, $C_g(\cdot)$ are affine, therefore the convex envelope of $C_g(\cdot)$ is equal to $C_g(\cdot)$ itself, as demonstrated in \cite{HuaCHP}. Given this condition, and considering that strong duality holds for SOCs \cite{andersen2002notes}, optimal dual values associated with system-wide constraints represent the dual maximisers of the Lagrangian dual problem, i.e. the convex hull prices.

\begin{figure}[!t]
    \centering
    \includegraphics[width=\columnwidth]{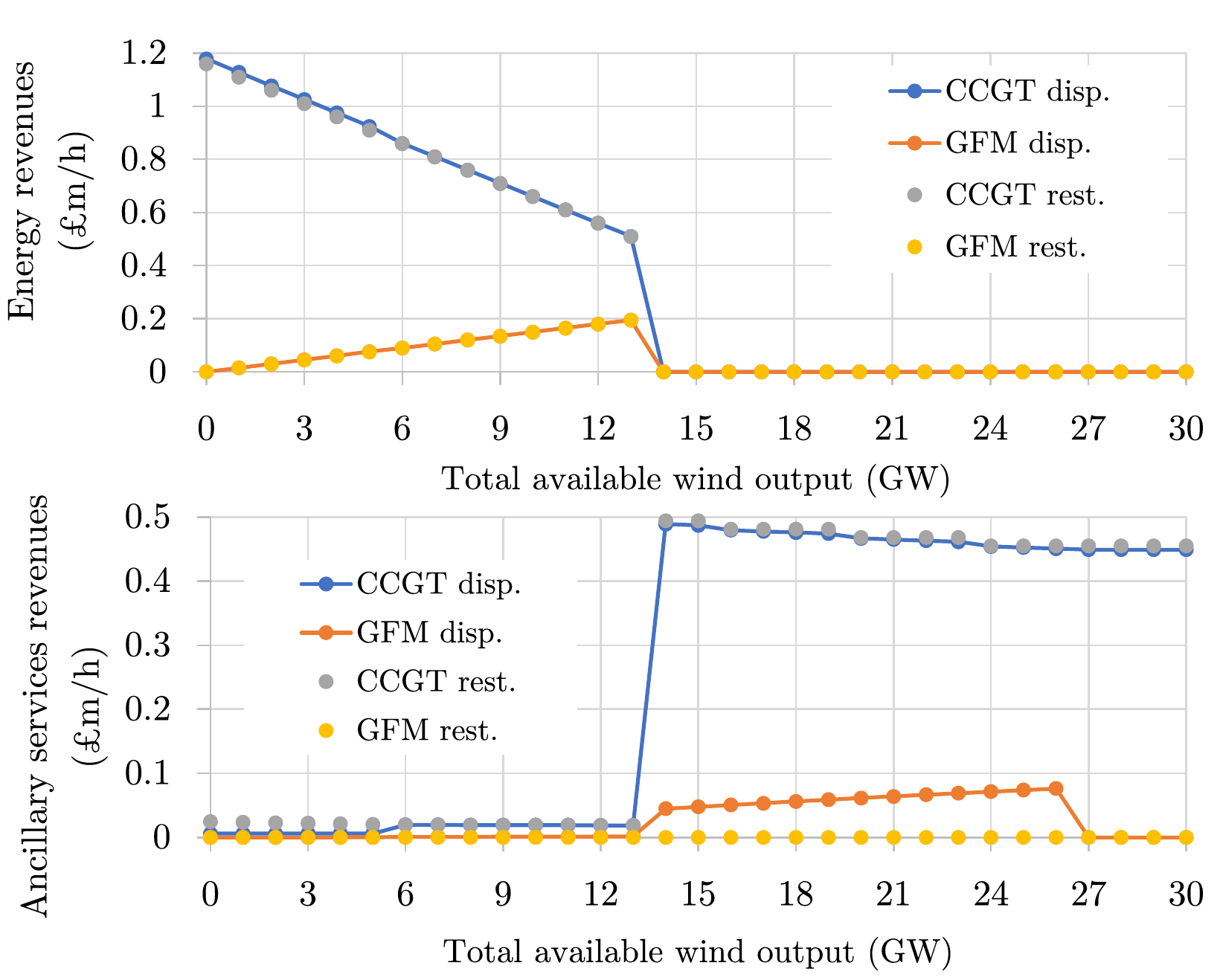}
    \caption{Hourly revenues for energy and ancillary services for all CCGTs and all wind GFM, depending on total wind output available in the system. Revenues computed through both dispatchable and restricted methods, where CHP are equivalent to dispatchable prices in this case.}
    \label{fig:Revenues_DispatRestr}
\end{figure}

In summary, the results for dispatchable prices shown in this section also correspond to CHP. However, computing CHP within more general frameworks is a complex issue which is still an active area of research.
Future work should study, among others, the implications of applying CHP for ancillary services while including ramp constraints for thermal generators, which would modify the mathematical definition of `$\textrm{conv}(\mathcal{X}_g)$' \cite{HuaCHP}, as well as including non-convexities within the system-wide constraints, such as the AC power flow equations \cite{Generalized_CHP}.

Now we show in Fig.~\ref{fig:Revenues_DispatRestr} the results from applying the different pricing methods to a case study with 30\% of the wind fleet providing synthetic inertia with a fixed $\textrm{H}_i=5\textrm{s}$. While the revenues from energy are roughly equal if computed with either the dispatchable or restricted pricing methodologies, these results show an important issue of the restricted pricing method: the revenues for ancillary services perceived by wind-GFM units are zero. 
CCGT units still perceive appropriate revenues associated to ancillary services, and they do so through the `commitment price', i.e.~the price associated to the dual variable $\lambda_\textrm{commit}$. In fact, the revenues from ancillary services for CCGTs are roughly equal in both methodologies, meaning that the price for commitment `absorbs' the prices for inertia and frequency response reserves. However, the value of all dual variables associated to the RoCoF, nadir and {q-s-s} constraints all drop to zero when applying the restricted pricing method.

In conclusion, restricted pricing is not suitable for placing incentives for RES to provide ancillary services. This is due to the fact that the price of ancillary services becomes integrated in the price of `commitment', and given that RES such as wind and solar do not have a commitment variable, the incentives for them to provide ancillary services would be removed if restricted pricing was used.
Therefore, convex hull pricing is suggested as the main method to be further studied for working in conjunction with the ancillary-services pricing proposed here. Additional pricing methodologies such as `incremental cost pricing' \cite{ONeillIncrementalPricing}
could also be considered.

\subsection{Pricing frequency-control services within multi-period UC}

Finally, to demonstrate the applicability of this pricing scheme to a multi-period Unit Commitment, the optimisation problem described in Section~\ref{sec:UC_formulation} is solved here for each of the 24 hours in a day. The objective function now becomes:
\begin{alignat}{3}
    & {\text{min}} \quad
    && \sum_{t=1}^{24} 
    \sum_{g \in \mathcal{G}} \textrm{c}_g^\textrm{st}\cdot y_g^\textrm{sg}(t) +
    \textrm{c}^\textrm{nl}_g\cdot y_g(t) + \textrm{c}_g^\textrm{m}\cdot P_g(t)
\end{alignat}
Where we consider a value of $\textrm{c}_g^\textrm{st} = \textrm{£} 10{,}000$. Along with all the constraints included in Section~\ref{sec:UC_formulation}, that apply to each hour independently, the following linking constraints for each hour are also included in the optimisation:
\begin{alignat}{2}
    & y_g(t) = y_g(t-1) + y_g^\textrm{sg}(t) - y_g^\textrm{sd}(t) &&  \label{eq:CommitmentTime}\\
    & y_g^\textrm{sg}(t) = y_g^\textrm{st}(t-\textrm{T}_g^\textrm{st}) &&  \label{eq:StartUpIndicator}\\
    & y_g^\textrm{st}(t) \leq
    \left[1-y_g(t-1)\right] - \hspace{-1mm} \sum_{j=t-\textrm{T}_g^\textrm{mdt}}^t y_g^\textrm{sd}(j) &&  \label{eq:StartUp}\\ 
    & y_g^\textrm{sd}(t) \leq
    y_g(t-1) - \hspace{-1mm} \sum_{j=t-\textrm{T}_g^\textrm{mut}}^ty_g^\textrm{sg}(j) &&  \label{eq:DownTime}
\end{alignat}
All the above constraints apply $\forall g \in \mathcal{G}, \; \forall t \in \{1, ..., 24\}$.

Constraints~(\ref{eq:CommitmentTime}) through (\ref{eq:DownTime}) model the operation of thermal units. Constraint~(\ref{eq:CommitmentTime}) defines the commitment state of each generator as `online' if the unit was already generating in the previous period (unless it has been shut down in the current period), or has started generating in the current period. Constraint~(\ref{eq:StartUpIndicator}) sets the state of a unit as `starts generating' in the current time period if the unit was started up `$\textrm{T}_g^\textrm{st}$' periods earlier, where we use a value of `$\textrm{T}_g^\textrm{st}=4\textrm{h}$' in this section. A unit can only be started up if it has been offline for at least `$\textrm{T}_g^\textrm{mdt}$' periods and was offline in the previous period, as defined by constraint~(\ref{eq:StartUp}). Finally, constraint~(\ref{eq:DownTime}) enforces that a generator can only be shut down if it was online in the previous period, as well as having been online for at least `$\textrm{T}_g^\textrm{mut}$' periods. Values of $\textrm{T}_g^\textrm{mdt}=1\textrm{h}$ and $\textrm{T}_g^\textrm{mut}=4\textrm{h}$ are used in the following case studies.

\begin{figure}[!t]
    \centering
    \includegraphics[width=\columnwidth]{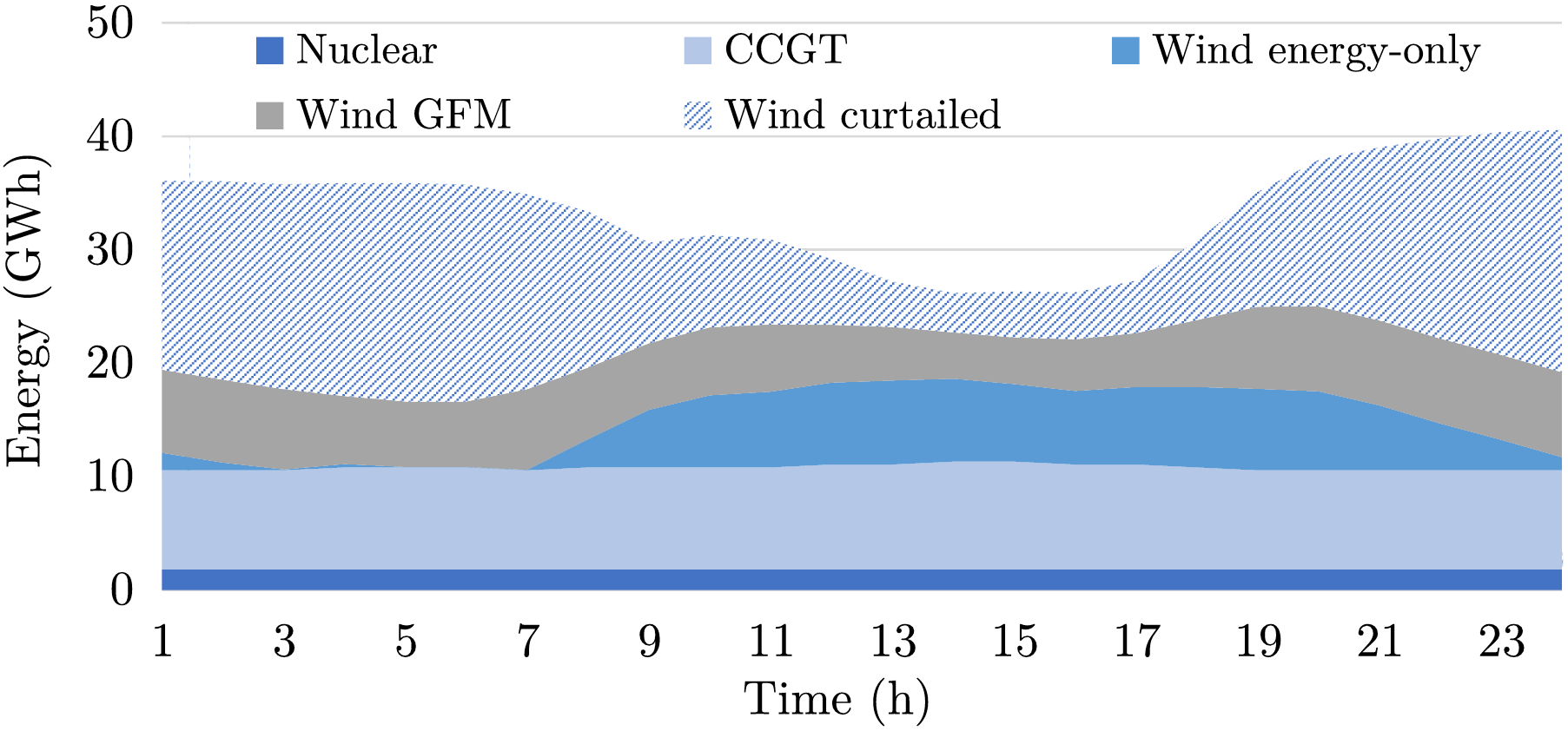}
    \caption{Energy delivered by each type of generation, for a 24h optimisation with 30GW of wind capacity, 30\% of which is GFM-equipped.}
    \label{fig:MultiPeriodGFM_Energy}
\end{figure}

\begin{figure}[!t]
    \centering
    \includegraphics[width=\columnwidth]{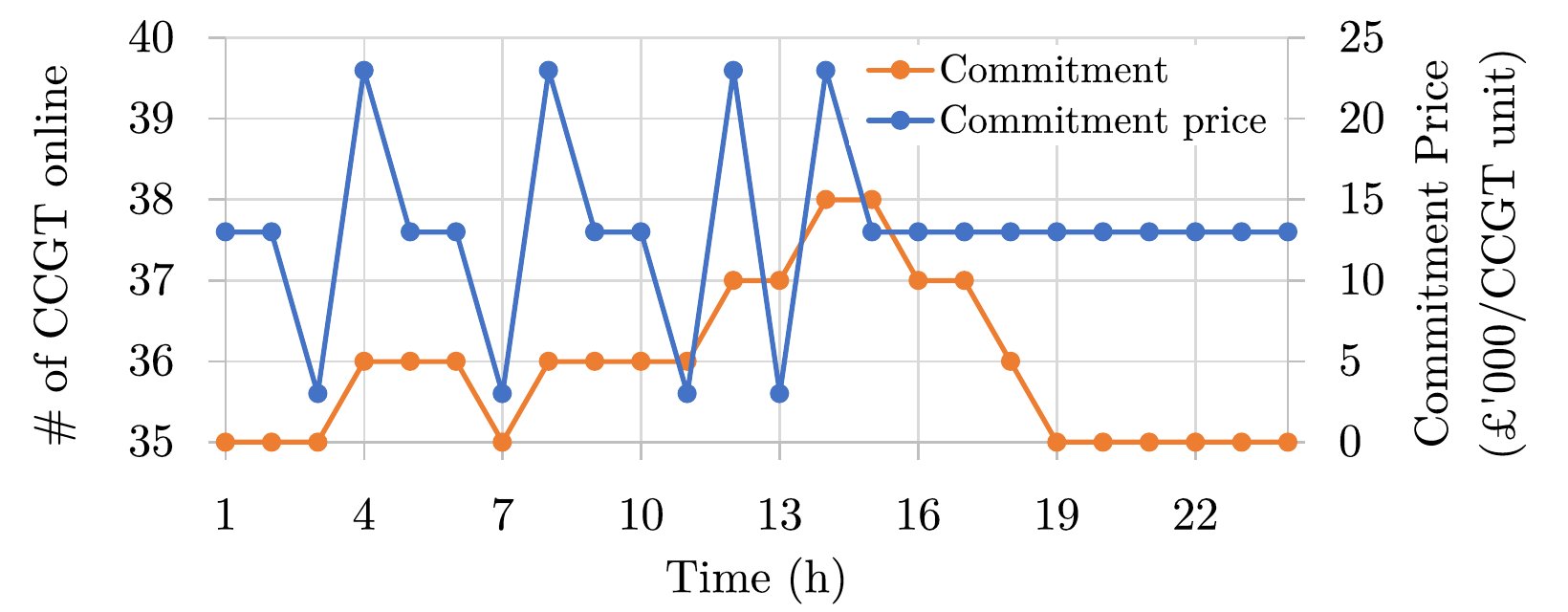}
    \caption{Number of CCGT units online in each hour and prices for commitment, computed through the restricted pricing method. From a 24h optimisation with 30GW of wind capacity, 30\% of which is GFM-equipped. }
    \label{fig:MultiPeriod_Commitment}
\end{figure}

Simulations are run here using a daily demand profile of 507GWh with a peak of 25GW occurring at 8pm, and 30GW of installed wind capacity. Both the dispatchable and restricted pricing methodologies are evaluated in this section. Regarding CHP, these are still equivalent to dispatchable prices for the multi-period UC considered here: additional non-convexities are introduced in this multi-period setting by binary variables $y_g^\textrm{st}$, $y_g^\textrm{sd}$ and $y_g^\textrm{sg}$, as well as start-up costs appearing in the individual cost functions for each generator; however, as demonstrated in \cite{HuaCHP}, all the mathematical deductions for CHP included in Section~\ref{subsection:sp_pricing} also apply for this case, given that strong duality still applies in our framework. 

\begin{figure}[!t]
    \centering
    \includegraphics[width=\columnwidth]{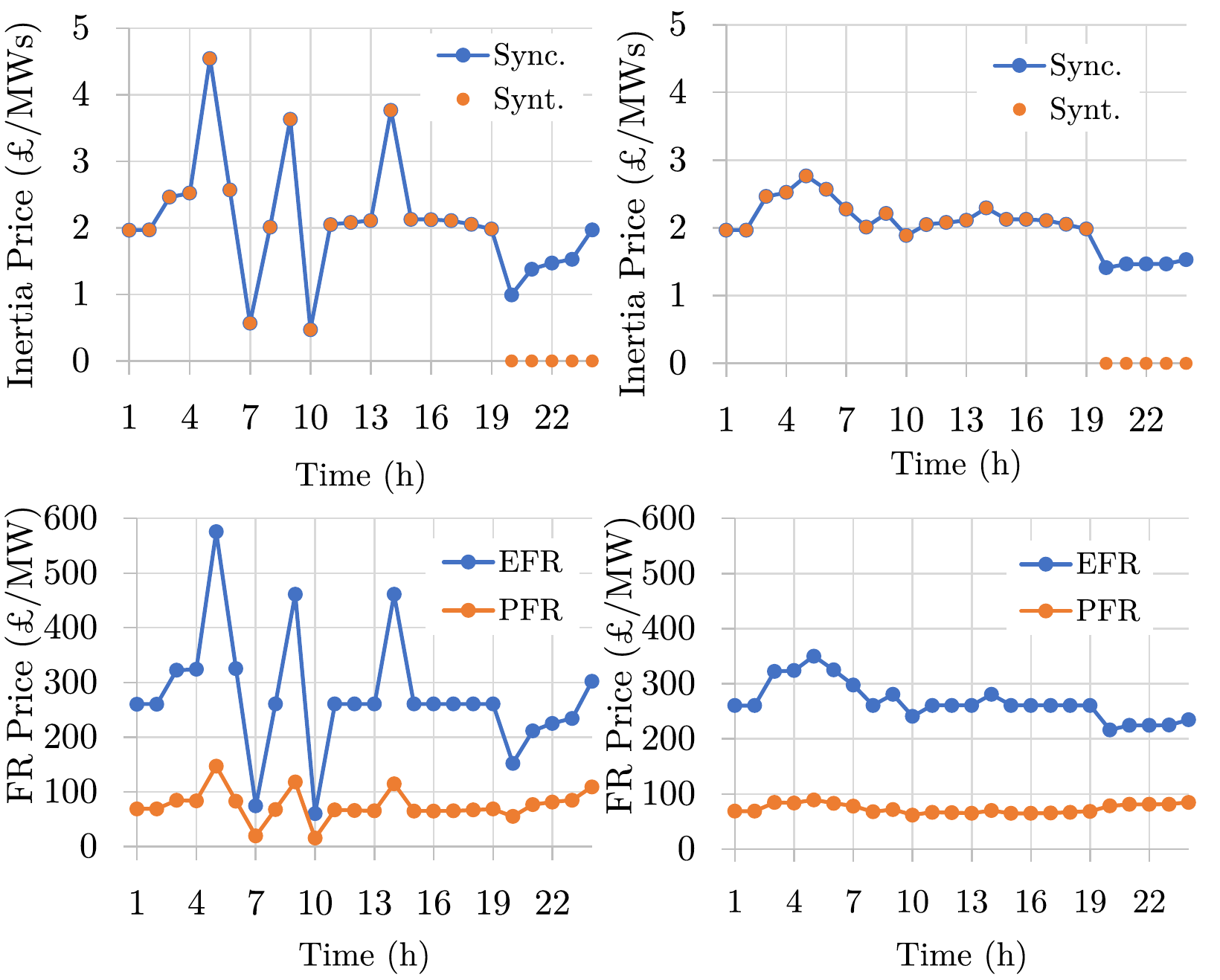}
    \caption{Prices for inertia and frequency response reserves, computed through the dispatchable pricing method. From a 24h optimisation with 30GW of wind capacity, 30\% of which is GFM-equipped. Left column is computed with $\textrm{c}_g^\textrm{st} = \textrm{£} 10{,}000$, right column with $\textrm{c}_g^\textrm{st} = \textrm{£} 1{,}000$. }
    \vspace{-4mm}
    \label{fig:MultiPeriodGFM}
\end{figure}

The results for a simulation with 30\% of the wind fleet equipped with GFM inverters and $\textrm{H}_i=5\textrm{s}$ are included in Figs.~\ref{fig:MultiPeriodGFM_Energy} to \ref{fig:MultiPeriodGFM}. The price for commitment shown in Fig.~\ref{fig:MultiPeriod_Commitment} takes on average a value of £13,000/h for each CCGT unit, which is equivalent to the revenues for inertia and PFR that a CCGT would perceive if dispatchable pricing were to be applied instead. On the other hand, as discussed in Section~\ref{subsection:sp_pricing}, the prices of inertia (both synchronous and synthetic) and frequency response reserves drop to zero when applying restricted pricing. The prices of these services computed using the dispatchable pricing method are shown in Fig.~\ref{fig:MultiPeriodGFM}, which contains some price spikes in certain hours. These spikes are related to the start-up cost of thermal units, as the price profile is flattened out if this cost is reduced from $\textrm{c}_g^\textrm{st} = \textrm{£} 10{,}000$ to $\textrm{£} 1{,}000$.

\begin{figure}[!t]
    \centering
    \includegraphics[width=\columnwidth]{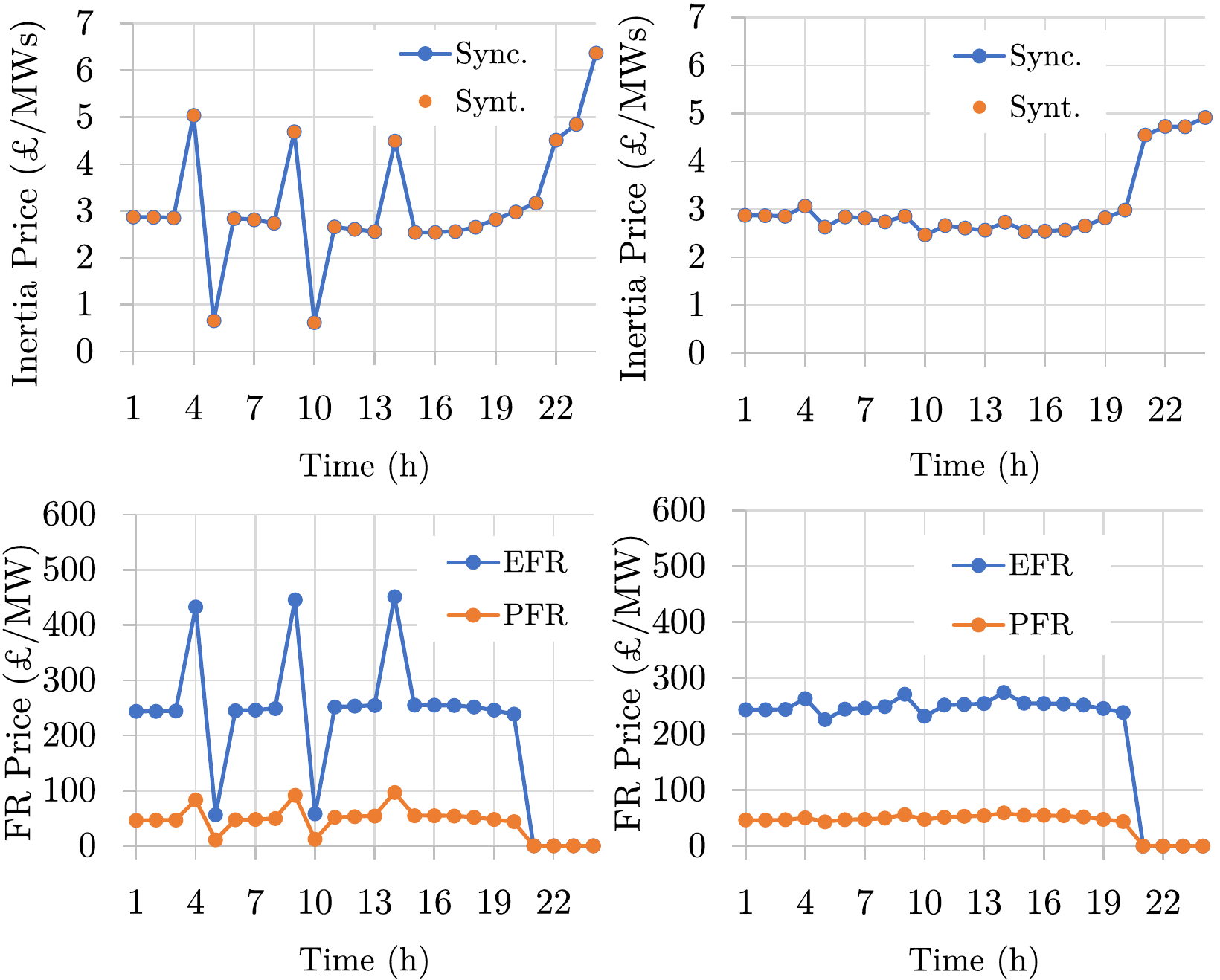}
    \caption{Prices for inertia and frequency response reserves, computed through the dispatchable pricing method. From a 24h optimisation with 30GW of wind capacity, 15\% of which can provide EFR. Left column is computed with $\textrm{c}_g^\textrm{st} = \textrm{£} 10{,}000$, right column with $\textrm{c}_g^\textrm{st} = \textrm{£} 1{,}000$. }
    \label{fig:MultiPeriodGFL}
\end{figure}

For completeness, the prices for frequency-control services for a case where 15\% of the wind fleet has the capability to provide EFR are shown in Fig.~\ref{fig:MultiPeriodGFL}. Similar price spikes can be observed, which would be smoothed out if the start-up costs were lower. Note that the price of frequency response reserves drops to zero for the last hours of the day in Fig.~\ref{fig:MultiPeriodGFL}, which coincides with the period of highest wind curtailment, and equivalently the price of synthetic inertia drops to zero in those same hours in Fig.~\ref{fig:MultiPeriodGFM}. This is the same behaviour that was shown in Sections~\ref{sec:WindEFR} and \ref{subsection:val_SI_GFM}, demonstrating that the prices of these services could drop to zero during times of very high RES output, due to an excess of either synthetic inertia or frequency response reserves delivered by RES.

Regarding the computational burden of this multi-period, frequency-secured UC problem, it was solved in under 1~second using Gurobi 9.5.0 in an eight-core 3.60 GHz Intel Core i7 CPU, with 64 GB of RAM. The problem contains 4,800 binary variables, 2,664 continuous variables and 20,530 constraints, out of which 24 are SOC constraints and the rest are linear. Since the problem is an MISOCP, scalability to bigger instances could be handled efficiently with off-the-shelf commercial software \cite{BoydConvex}.

\section{Conclusion and Future Work} \label{sec:Conclusion}

In this paper a shadow pricing formulation for ancillary services from RES and thermal generators has been proposed, compatible with a joint scheduling of energy and frequency-control services. This pricing scheme incentivises RES owners to enter the ancillary services market, and can potentially remove the need for any thermal generator to provide inertia. A combination of synthetic inertia (from GFM inverters) and frequency response from curtailed RES has been shown to achieve this goal, as synthetic inertia needs a slower but sustained source of power injection to compensate its recovery effect, therefore a certain level of RES curtailment is necessary. In addition, GFM assets have been shown to significantly increase their revenue if able to optimise the value of their synthetic inertia constant in every market settlement period, depending on overall RES power output in the wider grid.

As future work, it is important to further investigate the implications of the proposed pricing framework for ancillary services when applied within different methodologies for dealing with non-convexities, such as convex hull pricing. This is particularly relevant as this paper has shown that `restricted' pricing, i.e.~fixing the commitment decision for thermal units when computing prices, would remove the incentives for RES generators to participate in the market for frequency control services.

Another relevant area for future analysis is detailed modelling of the controls for GFM and GFL inverters, to work in tandem with the proposed pricing scheme. This would provide reassurance on the effectiveness of this method to guarantee frequency stability in real-time operation, while putting in place appropriate incentives for investment in the different technologies. Furthermore, other types of stability should be considered, such as voltage and transient stability, to define a market framework that makes full use of the capabilities of inverter-based resources and enables these to become the dominant assets in large-scale power grids. Finally, given that the cost of ancillary services can significantly increase in future decarbonised grids, it becomes relevant to understand who should bear these costs.

\section*{Acknowledgment}

The authors would like to thank the anonymous reviewers for their constructive criticism, which contributed to significantly enhance the quality of the paper and eventually led to new insights.

\appendices
\section{Deduction of nadir constraint} \label{ap:NadirConstraint}

To obtain the condition for keeping the frequency nadir above the limit that would trigger UFLS relays, i.e.~$\Delta f_\textrm{max}$, the swing equation must be solved for the instant when RoCoF becomes zero, that is, the frequency minimum is reached. First of all, an expression for the exact time when the nadir occurs, $t_\textrm{nadir}$, must be obtained:
\begin{equation} \label{eq:SwingAtNadir}
    \textrm{FR}(t_\textrm{nadir}) =  P_\textrm{L}
\end{equation}
For now we assume that $t_\textrm{nadir}>\textrm{T}_\textrm{EFR}=1\textrm{s}$, as otherwise there would not be sufficient inertia to secure the maximum RoCoF, eq.~(\ref{eq:Rocof}). Later on in this Appendix, we demonstrate why the opposite case, $t_\textrm{nadir}<\textrm{T}_\textrm{EFR}$, would not be possible in practice. Therefore, the expression for frequency response in eq.~(\ref{eq:SwingAtNadir}) is:
\begin{equation} \label{eq:FR_AtNadir}
    \textrm{FR}(t_\textrm{nadir}) = \sum_{ i \in \mathcal{I} } R_i + \sum_{ g \in \mathcal{G} } R_g \cdot \frac{ t_\textrm{nadir} } { \textrm{T}_\textrm{PFR} }
\end{equation}
Substituting (\ref{eq:FR_AtNadir}) into (\ref{eq:SwingAtNadir}):
\begin{equation}
    t_\textrm{nadir} = \frac{ \left( P_\textrm{L} - \sum_{ i \in \mathcal{I} } R_i \right) \cdot \textrm{T}_\textrm{PFR} }
    { \sum_{ g \in \mathcal{G} } R_g } 
\end{equation}

Now, the expression for `$\Delta f(t_\textrm{nadir})$' must be obtained by integrating the swing equation. The absolute value of this magnitude is:
\begin{multline} \label{eq:IntegralSwing}
    \left| \Delta f(t_\textrm{nadir}) \right| = \\
    \frac{f_0}{2(H_\textrm{sync} + H_\textrm{synt})} \cdot \int_0^{t_\textrm{nadir}}
    \left[ P_\textrm{L} -  {\textrm{FR}(t)} \right] \textrm{d} t
\end{multline}

By solving eq.~(\ref{eq:IntegralSwing}) and enforcing $\left| \Delta f(t_\textrm{nadir}) \right| \leq \Delta f_\textrm{max}$, constraint~(\ref{eq:nadir}) is finally obtained.

Note that for obtaining~(\ref{eq:nadir}), we have assumed above that $t_\textrm{nadir}>\textrm{T}_\textrm{EFR}$. Let us now consider the opposite case, i.e.~$t_\textrm{nadir}<\textrm{T}_\textrm{EFR}$. Following the same mathematical steps as above, the resulting nadir constraint would be:
\begin{equation} \label{NadirBeforeEFR}
    \frac{ H_\textrm{sync} + H_\textrm{synt} } {f_0}   
    \left( \frac{ R_\mathcal{I} }{ \textrm{T}_\textrm{EFR} } + \frac{ R_\mathcal{G} }{ \textrm{T}_\textrm{PFR} } \right)
    \geq 
    \frac{ \left( P_\textrm{L} \right) ^2 }{ 4 \Delta f_\textrm{max} } 
\end{equation}

To demonstrate that expression~(\ref{NadirBeforeEFR}) would require a very low value of system inertia, that would violate the RoCoF limit eq.~(\ref{eq:Rocof}), let us consider that the volume of PFR available is zero (if any PFR was available, an even lower volume of inertia would be required). In this case, the volume of EFR needed to reach the nadir equilibrium would be $P_\textrm{L}$, therefore (\ref{NadirBeforeEFR}) becomes:
\begin{equation} 
    \frac{ H_\textrm{sync} + H_\textrm{synt} } {f_0}   
    \cdot \frac{ P_\textrm{L} }{ \textrm{T}_\textrm{EFR} } 
    \geq 
    \frac{ \left( P_\textrm{L} \right) ^2 }{ 4 \Delta f_\textrm{max} } 
\end{equation}
Then, the level of inertia required for the above constraint to be binding is:
\begin{equation} 
    H_\textrm{sync} + H_\textrm{synt}  
    =
    \frac{  P_\textrm{L}  \cdot f_0 \cdot \textrm{T}_\textrm{EFR}}{ 4 \Delta f_\textrm{max} } 
\end{equation}
Substituting now this value into the RoCoF constraint, expression~(\ref{eq:Rocof}), we obtain:
\begin{equation} 
    \textrm{T}_\textrm{EFR}  
    \geq
    \frac{  2 \Delta f_\textrm{max} }{ \textrm{RoCoF}_\textrm{max} } 
\end{equation}

Only if the above condition holds, could the nadir constraint~(\ref{NadirBeforeEFR}) be binding while also respecting the RoCoF limit defined by~(\ref{eq:Rocof}). For the case of Great Britain considered in the case studies in Section~\ref{sec:CaseStudies}, the values of $\textrm{RoCoF}_\textrm{max}=1\textrm{Hz/s}$ and $\Delta f_\textrm{max}=0.8\textrm{Hz}$ would require a value of $\textrm{T}_\textrm{EFR} \geq 1.6\textrm{s}$, which is inconsistent with the value of $\textrm{T}_\textrm{EFR}=1\textrm{s}$ required in GB.

\ifCLASSOPTIONcaptionsoff
  \newpage
\fi

\IEEEtriggeratref{23}

\bibliographystyle{IEEEtran} 
\bibliography{Luis_PhD}

\vskip -1\baselineskip plus -1fil

\begin{IEEEbiographynophoto}{Luis Badesa}
(S'14-M'20) received the PhD degree in Electrical Engineering from Imperial College London, U.K., in 2020. He will start as Assistant Professor in Electrical Engineering at the Technical University of Madrid (UPM), Spain, and is currently a Research Associate at Imperial College London. His research focus is on modelling the operation and economics of low-inertia electricity grids, and market design for frequency-containment services.
\end{IEEEbiographynophoto}

\vskip -1\baselineskip plus -1fil

\begin{IEEEbiographynophoto}{Carlos Matamala}
(S'22) is Electrical Engineer from the University of Chile (UCH). He worked at the Energy Center UCH for five years, doing research and consultancy to support decision-makers in the energy sector. Currently, he is pursuing a PhD in Electrical Engineering at Imperial College London, U.K. His research interests include operation, planning, and market design for low-carbon energy systems.
\end{IEEEbiographynophoto}

\vskip -1\baselineskip plus -1fil

\begin{IEEEbiographynophoto}{Yujing Zhou}
received the BEng in Energy Science and Engineering from the City University of Hong Kong, China, in 2020, and the MSc in Future Power Networks from Imperial College London, U.K, in 2021.
She is currently an Electrical Engineer within the building services industry. Her research is focused on sustainable energy utilisation and future energy markets.
\end{IEEEbiographynophoto}

\vskip -1\baselineskip plus -1fil

\begin{IEEEbiographynophoto}{Goran Strbac}
(M'95) is Professor and Chair in Electrical Energy Systems at Imperial College London, U.K. His current research is focused on the optimisation of operation and investment of low-carbon energy systems, energy infrastructure reliability and future energy markets. 
\end{IEEEbiographynophoto}

\end{document}